# Brightening of Long, Polymer-Wrapped Carbon Nanotubes by *sp³* Functionalization in Organic Solvents

*Felix J. Berger†,‡, Jan Lüttgens†,‡, Tim Nowack †, Tobias Kutsch †,§, Sebastian Lindenthal†, Lucas Kistner†, Christine C. Müller†, Lukas M. Bongartz †, Victoria A. Lumsargis †,#, Yuriy Zakharko†,⊥ and Jana Zaumseil\*†,‡*

†Institute for Physical Chemistry, Universität Heidelberg, D-69120 Heidelberg, Germany

‡Centre for Advanced Materials, Universität Heidelberg, D-69120 Heidelberg, Germany

§Institute of Physical Chemistry, RWTH Aachen University, D-52074 Aachen, Germany

**Corresponding Author**

\*zaumseil@uni-heidelberg.de

**Present Addresses**

# Department of Chemistry, Purdue University, West Lafayette, IN 47907, USA

⊥ Institute of Semiconductor and Solid State Physics, Johannes Kepler University Linz, A-4040 Linz, Austria

**ABSTRACT**

The functionalization of semiconducting single-walled carbon nanotubes (SWNTs) with $sp^3$ defects that act as luminescent exciton traps is a powerful means to enhance their photoluminescence quantum yield (PLQY) and to add optical properties. However, the synthetic methods employed to introduce these defects are so far limited to aqueous dispersions of surfactant-coated SWNTs, often with short tube lengths, residual metallic nanotubes and poor film formation properties. In contrast to that, dispersions of polymer-wrapped SWNTs in organic solvents feature unrivaled purity, higher PLQY and are easily processed into thin films for device applications. Here, we introduce a simple and scalable phase-transfer method to solubilize diazonium salts in organic nonhalogenated solvents for the controlled reaction with polymer-wrapped SWNTs to create luminescent aryl defects. Absolute PLQY measurements are applied to reliably quantify the defect-induced brightening. The optimization of defect density and trap depth results in PLQYs of up to 4 % with 90 % of photons emitted through the defect channel. We further reveal the strong impact of initial SWNT quality and length on the relative brightening by $sp^3$ defects. The efficient and simple production of large quantities of defect-tailored polymer-sorted SWNTs enables aerosol-jet printing and spin-coating of thin films with bright and nearly reabsorption-free defect emission, which are desired for carbon nanotube-based near-infrared light-emitting devices.

***Keywords:*** single-walled carbon nanotubes, $sp^3$ defects, functionalization, photoluminescence, diazonium salts

The combination of narrowband emission in the near-infrared (NIR) and high charge carrier mobilities make semiconducting single-walled carbon nanotubes (SWNTs) highly desirable materials for optoelectronic devices, such as light-emitting diodes[1, 2] and field-effect transistors.[3-5] The fabrication of such devices based on homogeneous thin films requires concentrated SWNT inks without any metallic nanotubes[6] that can be processed easily and reproducibly. While photoluminescence (PL) from SWNTs was first observed in aqueous dispersions stabilized by surfactants,[7] nowadays selective wrapping with conjugated polymers in organic solvents is often applied for the extraction of semiconducting SWNTs with the highest purity and photoluminescence quantum yields (PLQYs).[8] In conjunction with mild and scalable exfoliation methods such as shear force mixing,[9] large volumes of dispersions with high concentrations of long SWNTs with high quality can be produced by polymer-wrapping. These dispersions are ideal for the deposition of homogeneous films by spin-coating,[10] aerosol-jet[11] or inkjet printing,[12] owing to the beneficial properties of organic solvents (surface tension, viscosity, vapor pressure *etc.*). However, despite the substantial progress in terms of purification and processing, light-emitting devices based on carbon nanotubes are still limited by their low absolute PLQYs (~1 % in dispersion and ~0.1 % in films) resulting from the presence of low-lying dark exciton states[13] and the fast diffusion of mobile excitons to non-radiative quenching sites.[14, 15]

The most promising approach to enhance the PLQY of SWNTs is their functionalization with a limited number of oxygen[16, 17] or *sp³* defects[18-20] that act as luminescent exciton traps with depths on the order of 100 meV. Excitons trapped at these defect sites are not subject to otherwise dominant diffusion-limited contact-quenching,[14] and the modification of the local electronic structure by the defects enables radiative relaxation and red-shifted emission (labelled as $E_{11}^*$, or $E_{11}^{*-}$ for different binding configurations).[21-24] As demonstrated by Piao *et al.*,[18] the emission from *sp³* defects that are introduced to the SWNT lattice *via* aryldiazonium chemistry can be much brighter than that of pristine SWNTs. This report motivated a series of studies on the population mechanism[25] and relaxation dynamics[26-29] of these emissive trap states and the impact of the different binding configurations on their optical properties.[22, 30-32] Moreover, it was shown that $sp^3$ defects in SWNTs can be employed for room temperature single photon emission in the NIR,[29-31, 33] PL imaging within the second biological window[34] and sensing.[35] However, the synthetic methods for the creation of such $sp^3$ defects usually involve the use of highly polar reagents,[18, 36] such as diazonium salts. They are thus limited to dispersions of SWNTs in water. While it is possible to sort semiconducting SWNTs by chromatography or two-phase extraction of aqueous dispersions,[37, 38] the superior purity of polymer-sorted SWNTs for a minimum of purification steps makes them preferable for optoelectronic applications.[6, 9, 39, 40]

Given the advantages of polymer-wrapped SWNTs in organic solvents it is highly desirable to develop an easy and reproducible $sp^3$ functionalization method for these systems. The main challenge is the conflict between the solubility of polar diazonium salts and the colloidal stability of SWNTs wrapped with nonpolar conjugated polymers. Although the *in situ* generation of diazonium salts may be used, this approach suffers from limited scalability as it requires elevated temperatures and inert conditions.[41, 42] Other attempts, for example, by dip-doping were limited to pre-deposited polymer-sorted nanotubes and only successful with very reactive diazonium salts.[31]

Here, we present a simple and scalable method to create luminescent $sp^3$ defects in polymer-wrapped (6,5) SWNTs in organic nonhalogenated solvents. The addition of a polar co-solvent and a phase-transfer agent allows the preformed diazonium salts to be solubilized and to react with the SWNTs at room temperature. Due to the high concentration and purity of the resulting dispersions of functionalized SWNTs, the absorption features arising from the defects can be analyzed in detail for the first time. These data give insights into reorganization processes upon exciton trapping and site-to-site interactions. To reliably quantify the functionalization-induced brightening, we determine the PLQY by absolute measurements in an integrating sphere and demonstrate the influence of defect density and trap depth on the PLQY enhancement. We reveal the strong impact of initial SWNT quality and length on the relative brightening by $sp^3$ defects. This simple functionalization method provides access to large quantities of polymer-wrapped and $sp^3$ functionalized SWNTs that can be easily processed into thin or thick films with enhanced PLQY that may enable brighter NIR light-emitting devices.

**RESULTS AND DISCUSSION**

Our method for creating luminescent $sp^3$ defects in polymer-wrapped SWNTs relies on the solubilization of preformed aryldiazonium salts by an ether crown[43] (here, 18-crown-6) in a moderately polar medium (see **Figure 1**). This reaction approach avoids chlorinated solvents while facilitating large scale functionalization of polymer-wrapped carbon nanotubes with a minimum of processing steps. As a model system we use (6,5) SWNTs, which are extracted with high purity by selective dispersion with the polyfluorene copolymer PFO-BPy in toluene (see **Methods**). A mixture of toluene and acetonitrile (MeCN) in an 80:20 vol-% ratio is sufficiently polar to solubilize aryldiazonium salts at concentrations on the order of a few g L$^{-1}$ in the presence of 18-crown-6 as a phase-transfer agent, yet does not lead to substantial aggregation of (6,5) SWNTs or free PFO-BPy as corroborated by absorption and PL spectra (see the **Supporting Information, Figure S1**). A low concentration ($10^{-9}$ mol L$^{-1}$) of potassium acetate (KOAc) is included to optimize the selectivity of the functionalization process (for details see the **Supporting Information, Figure S2**). After a typical reaction time of 16 hours at room temperature and in the dark, the SWNTs are collected by vacuum filtration and unreacted diazonium salt is washed off with acetonitrile and toluene. This washing step is crucial as we found the SWNT PL emission of the reaction mixture to be strongly quenched by diazonium salts adsorbed on the SWNT surface. In aqueous media less than 10 % of the added diazonium salt actually reacts with the SWNT lattice to form *sp³* defects[44] and the reactivity in organic solvents turns out to be even lower. This drop in reactivity might result from differences in the reaction mechanism (see the **Supporting Information**), the surface accessibility provided by the wrapping polymer or steric hindrance due to ether crown complexation. Hence, it is likely that a large number of unreacted aryldiazonium cations are adsorbed on the SWNTs and quench the PL. After washing, the *sp³* functionalized (6,5) SWNTs are easily redispersed in pure solvent (here, toluene) by bath sonication and their intrinsic spectroscopic properties in dispersion can be studied. Note that the filtration and washing step does not constitute an additional workload as it also removes most of the excess polymer, which is often a necessary step in film and device fabrication. A detailed experimental protocol for *sp³* functionalization of PFO-BPy-wrapped (6,5) SWNTs is provided in the **Supporting Information**.

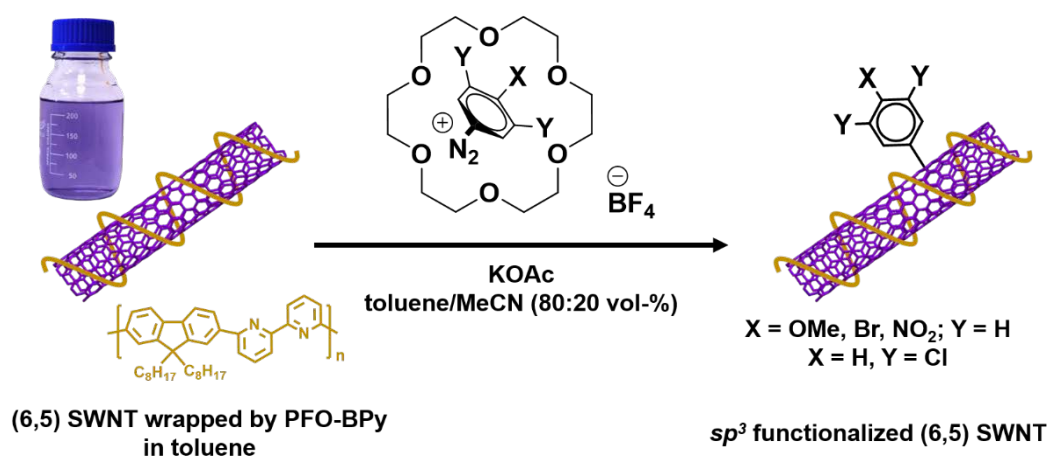

**Figure 1.** Schematic functionalization of PFO-BPy-wrapped (6,5) SWNTs by aryldiazonium salts in a toluene/acetonitrile mixture containing 18-crown-6 as a phase-transfer agent.

The overall procedure is scalable to large amounts of functionalized (6,5) SWNTs that can be redispersed at different low or high concentrations thus enabling reliable absorption measurements, determination of the absolute PLQY and creation of thick luminescent films. To establish the robustness and applicability of our procedure for polymer-wrapped SWNTs, we employ the outlined routine to introduce aryl defects from low to high defect densities and with substituents ranging from electron-withdrawing to electron-donating character, and then characterize their optical properties in detail.

First, we discuss the spectral characteristics of $sp^3$ functionalized PFO-BPy-wrapped (6,5) SWNTs using the example of 4-bromophenyl defects as shown in **Figure 2**. In addition to the strong $E_{11}$ transition at ~994 nm, the absorption spectra of the dispersions in **Figure 2a** show an absorption band at ~1142 nm that grows systematically with the concentration of reagent used. Since unreacted diazonium salt and adsorbed by-products were removed by washing and the untreated reference sample does not show any absorption features in this spectral range, we can unambiguously assign this band to $sp^3$ defects. Previous assignments of this transition in the literature were complicated by the presence of minority chiralities and low concentrations of functionalized SWNTs.[18] These issues are absent from dispersions prepared by polymer-wrapping. The highest $E_{11}^*/E_{11}$ absorbance ratio in this study was 0.13, however without knowledge of the extinction coefficient of the defect absorption this value still does not enable an absolute quantification of the number of defects in the lattice (*vide infra*).

**Emission Spectra and Stokes Shift.** Photoluminescence excitation at the $E_{22}$ transition (575 nm) of (6,5) SWNTs gave the emission spectra presented in **Figure 2b** showing PL arising from mobile ($E_{11}$) and trapped ($E_{11}^*$) excitons at ~999 nm and ~1161 nm, respectively. At the highest shown defect density, there is an additional red tail that stretches up to 1450 nm and has been attributed to deeper trap states ($E_{11}^{*-}$) resulting from different arrangements of aryl moieties on the SWNT lattice.[22, 30-32] Note that all emission features are red-shifted compared to functionalized (6,5) SWNTs dispersed in aqueous surfactant solutions owing to interaction with the conjugated wrapping polymer and differences in the dielectric environment.[45] Since the PL was collected through an objective, the NIR absorption of toluene does not affect the spectra as a result of the extremely short path length within the liquid. As expected, the $E_{11}*$ PL intensity increases relative to the $E_{11}$ with rising diazonium concentration, and thus defect density. Below a diazonium salt concentration of ~0.037 mmol L$^{-1}$, the $E_{11}$ and $E_{11}*$ transition energies depend only weakly on defect density, supporting the picture of well-isolated *sp$^3$* defects. However, above this level, the $E_{11}$ transition blue-shifts both in absorption and in emission with further increasing defect density, whereas the $E_{11}*$ transition red-shifts. The origin of these opposing spectral shifts is currently not well-understood, but may provide an interesting starting point for future computational studies. The spectral shifts are more pronounced in absorption than in emission and this difference is reflected clearly in the Stokes shifts (see **Figure 2c**). The untreated (6,5) SWNTs exhibit an $E_{11}$ Stokes shift of 6 meV that reaches 11 meV for highly functionalized samples. Conversely, the $E_{11}*$ Stokes shift drops from 18 meV to 14 meV as the defect density increases.

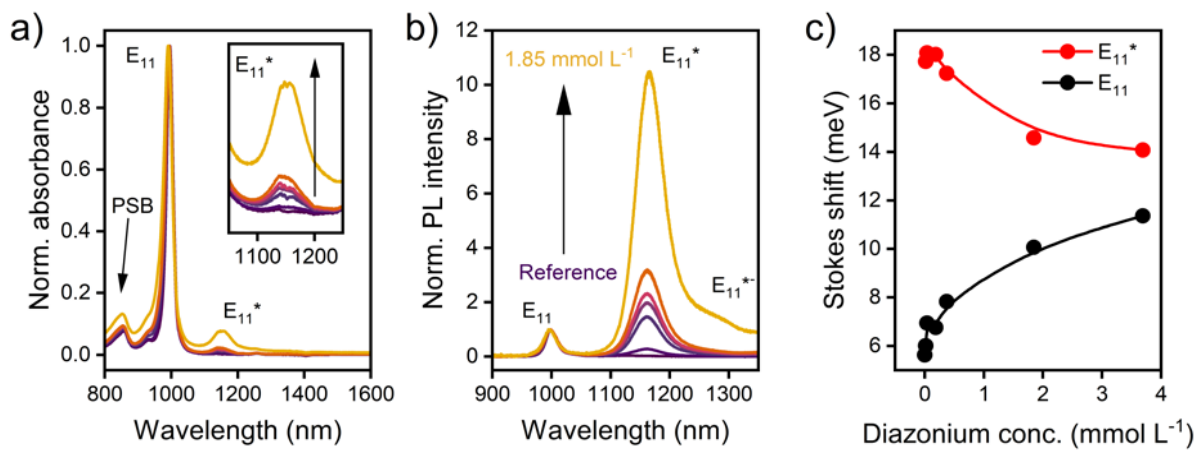

**Figure 2.** a) Normalized absorption spectra and b) PL spectra recorded under pulsed excitation (575 nm, ~0.5 mJ cm$^{-2}$) of PFO-BPy-wrapped (6,5) SWNTs in toluene functionalized using different concentrations of 4-bromobenzenediazonium tetrafluoroborate in the reaction mixture to adjust the defect density. c) Stokes shift of $E_{11}$ (black) and $E_{11}$* (red) transition as a function of diazonium salt concentration. Solid lines are guides to the eye.

The small $E_{11}$ Stokes shift originates from the rigidity of the SWNT structure (*i.e.*, small reorganization energy) in combination with the delocalized nature of the $E_{11}$ exciton.[46] In contrast to that, quantum chemical calculations predict a significant distortion in the vicinity of an *sp³* defect upon exciton trapping.[21, 26, 47, 48] At low defect densities the observed $E_{11}$* Stokes shift is about three times larger than that associated with the $E_{11}$ exciton. Generally, we find Stokes shifts of 18 to 21 meV depending on the substituent on the aryl group (see **Supporting Information, Figure S3**). Following the assignment of the $E_{11}$* transition in (6,5) SWNTs to the *ortho*$^{++}$ defect binding configuration[30] (equivalent to *ortho* L$_{90}$ in Ref. 22), the experimentally measured $E_{11}$* Stokes shift for 4-Br-phenyl defects is almost an order of magnitude smaller than predicted by calculations (18 *vs.* ~140 meV).[22] Even though part of this discrepancy might result from imperfect energy scaling in the quantum chemical calculations,[49] it is unlikely to be the sole origin. Further, the range of reorganization energies (20 to 120 meV) extracted from temperature-dependent PL measurements *via* a model involving vibrational reorganization upon exciton trapping is well above our measured $E_{11}$* Stokes shifts.[47] This mismatch suggests that the relationship between the optical trap depth given by the energy difference between the $E_{11}$ and $E_{11}$* transition and the thermal detrapping energy might be more complex than previously assumed. Nevertheless, the general trend in $E_{11}$* Stokes shifts reflects the reported behavior of reorganization energies.[47] Both quantities decrease as a function of defect density, most likely due to site-to-site interactions and associated electronic state delocalization.[47] **Figure S4** (**Supporting Information**) summarizes the PL spectra obtained for (6,5) SWNTs functionalized with other aryl defects bearing more electron-withdrawing (4-nitro and 3,5-dichloro) or electron-donating (4-methoxy) substituents. Note that all PL spectra are normalized to the $E_{11}$ intensity and thus no conclusions about the relative PLQYs can be drawn from these spectra alone.

**Defect Density Metrics.** An important experimental consideration is whether the density of the created defects scales linearly with the reagent concentration. Hence, we recorded Raman spectra of the functionalized SWNTs and determined the intensity of the D mode, which is expected to be proportional to the density of symmetry-breaking $sp^3$ defects,[50] relative to the $G^+$ mode, which is a measure of the density of $sp^2$ carbon atoms in the probed spot (**Supporting Information, Figure S5**). Similar to the functionalization process in an aqueous environment,[18] we find a roughly linear relationship between the D/$G^+$ ratio and the diazonium concentration (**Figure 3a**). In addition to this linearity, we observe that the reagent concentration window across which defect luminescence is detectable in organic solvent spans three orders of magnitude. In water this range is only one order of magnitude and occurs at much lower concentrations.[18] While this aspect is simply the result of the lower reactivity of the diazonium species in organic media, it is highly beneficial in practical terms, as it facilitates precise tuning of the defect density.

In addition to the D/G$^+$ ratio of the Raman signals, the ratio of E$_{11}$* to E$_{11}$ absorption was evaluated after background subtraction[51, 52] and plotted as a function of the diazonium concentration (**Figure 3a**). The correlation is roughly linear at low defect levels, but becomes sublinear at high levels. This dependence might be explained by assuming the oscillator strengths (per carbon atom) of pristine and functionalized regions in a 1D lattice to be constant at low defect densities. In this case, defect creation reduces the number of carbon atoms contributing to E$_{11}$ absorption, whereas E$_{11}$* absorption increases by an increment. Then, the E$_{11}$*/E$_{11}$ absorbance ratio is given by $\frac{f_d}{f_p} \frac{N_d}{N-N_d}$, with $f_d$ and $f_p$ as the oscillator strengths of defective and pristine lattice regions, respectively, $N_d$ denotes the number of *sp$^3$* hybridized carbon atoms and $N$ the total number of carbon atoms in the lattice. For low defect densities, *i.e.*, $N_d \ll N$, this geometric model predicts an approximately linear dependence in agreement with the experimental data. However, at high defect densities, a superlinear increase would be expected, which contradicts the measured sublinear behavior. This difference is likely caused by a drop in oscillator strength of the defects once they start clustering. Computational studies have already revealed the strong impact of the addition pattern around the aryl moiety on the oscillator strength of optical transitions.[21, 22] Furthermore, this trend is in agreement with the limiting case of a fully functionalized SWNT, for which all transitions vanish and only a scattering background remains.[53]

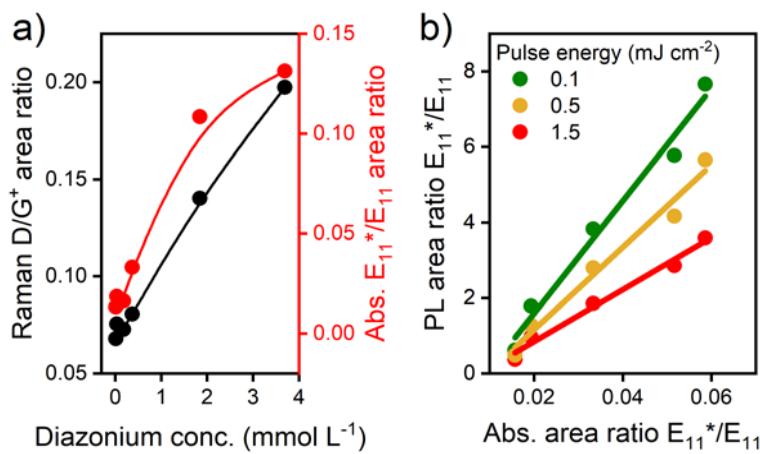

**Figure 3.** Raman, absorption and PL properties of functionalized (6,5) SWNTs using different concentrations of 4-bromobenzenediazonium tetrafluoroborate. a) Integrated Raman D/G$^+$ ratio and integrated $E_{11}^*/E_{11}$ absorbance ratio *versus* diazonium salt concentration. Solid lines are guides to the eye. b) Integrated $E_{11}^*/E_{11}$ PL ratio *versus* integrated $E_{11}^*/E_{11}$ absorbance ratio for different excitation densities and linear fits to the data.

Based on these different metrics for the *sp³* functionalization level, it is possible to test their relation to the emission properties of (6,5) SWNTs, in particular the integrated $E_{11}^*/E_{11}$ PL ratio. This ratio is plotted as a function of the diazonium concentration, the Raman D/G$^+$ ratio and the $E_{11}^*/E_{11}$ absorbance ratio in the **Supporting Information, Figure S6**. All graphs show a roughly linear correlation with the best fit found for the $E_{11}^*/E_{11}$ PL ratio *versus* the $E_{11}^*/E_{11}$ absorbance ratio. Importantly, even though the $E_{11}^*/E_{11}$ PL ratio is frequently used itself as a measure of defect density,[27, 47] it is only meaningful for a given excitation power density owing to the nonlinear behavior of both $E_{11}$ and $E_{11}^*$ emission. While PL from mobile excitons saturates at high excitation power due to exciton-exciton annihilation,[54, 55] the saturation of defect emission has been attributed to state-filling due to the fast exciton diffusion to trap sites (~10 ps) and the long lifetimes of localized excitons (~200 ps).[56] Since defect emission generally saturates at lower excitation power than $E_{11}$ emission, the $E_{11}^*/E_{11}$ PL ratio decreases as a function of pump power (see **Supporting Information, Figure S7a and S7b**). As shown in **Figure S7c**, PL spectra collected under lamp excitation (~1 mW cm$^{-2}$) are strongly dominated by defect emission. The impact of the power-dependence on the correlation between $E_{11}^*/E_{11}$ PL ratio and $E_{11}^*/E_{11}$ absorbance ratio can be seen in **Figure 3b** where the slope of the linear fit changes with excitation power.

In summary, the Raman D/G$^+$ ratio and the $E_{11}^*/E_{11}$ absorbance ratio can be used as metrics for the *sp*$^3$ defect density of functionalized polymer-wrapped SWNTs. The $E_{11}^*/E_{11}$ PL ratio should only be used as a measure of functionalization for a known and fixed excitation power and care should be take when comparing data from different experiments.

**Photoluminescence Quantum Yields and Lifetimes.** One of the main reasons to create luminescent *sp*$^3$ defects in SWNTs on a large scale is the goal to substantially increase their typically low PL quantum yields, *i.e.*, the ratio of emitted to absorbed photons, for practical applications. In the following we investigate the effect of *sp*$^3$ defects on the PL quantum yield of polymer-wrapped (6,5) SWNTs. At this point, it must be emphasized that the functionalization level that maximizes the PLQY of the ensemble of functionalized SWNTs is not necessarily the one which leads to the most efficient radiative relaxation at an individual trap site. The efficiency of exciton trapping and competition with diffusive contact-quenching are equally important. In other words, even if very low defect densities on the order of ~1 defect per nanotube support radiative relaxation of trapped excitons with the highest efficiency (*i.e.*, emitted photons per trapped excitons),[28, 31] they will not be able to compete with diffusive contact-quenching given the estimated quenching site densities on SWNTs of ~9 μm$^{-1}$.[57]

Here, we aim to understand how the PLQY of an ensemble of functionalized SWNTs is affected by the defect density, the substituent on the aryl group influencing the trap depth and the dispersion method, *i.e.*, quality and length of the SWNTs. To address these points without the uncertainties associated with the use of a reference fluorophore,[58] the PLQY was determined directly by measuring the laser absorption at the $E_{22}$ transition and PL emission of the sample in an integrating sphere (absolute method as demonstrated previously).[9] Since the PLQY, especially of the defect state, is a function of the excitation power density due to the nonlinear behavior discussed above, we note that all measurements were performed under pulsed excitation (pulse width ~6 ps) with an energy density of ~1.5 nJ cm$^{-2}$. As discussed in the **Supporting Information**, interaction between multiple excitons can be safely neglected at these extremely low pulse energies.

**Figure 4a** shows the evolution of the total PLQY as well as the spectral contributions of the main peaks, $E_{11}$ and $E_{11}$*, for shear-mixed (6,5) SWNTs functionalized with 4-nitrophenyl defects. Corresponding graphs for other substituents can be found in the **Supporting Information, Figure S8**. Note that unlike SWNTs in aqueous dispersions, polymer-wrapped (6,5) SWNTs already exhibit quite good PLQYs of up to 2.4 %.[9] After functionalization, even low defect densities lead to a sharp drop of $E_{11}$ PLQY and the emergence of strong $E_{11}$* emission. At 0.037 mmol L$^{-1}$, a local maximum in PLQY *vs.* diazonium concentration is observed, at which the $E_{11}$ still has an efficiency of 1.1 %, but the $E_{11}$* already contributes an additional 2.5 %. This range of defect density could be favorable when elevated PLQYs are needed, but trap sites should be well-isolated. We note that this maximum might be less pronounced for SWNTs with higher initial quenching site density and thus, lower starting levels of $E_{11}$ PLQY.

At a diazonium salt concentration of 0.37 mmol $L^{-1}$, the PLQY of the defect emission peaks at 3.5 % with a total PLQY of 3.8 %. Hence, more than 90 % of photons are emitted through the defect channel. This scenario leads to the highest $E_{11}^*$ and total PLQY observed in this concentration series. As the number of defects rises further, the $E_{11}$ contribution converges to zero, but the $E_{11}^*$ emission efficiency decreases as well. This observation is in agreement with the negligible PL found for highly functionalized SWNTs.[59] The outlined PLQY evolution of functionalized, polymer-wrapped (6,5) SWNTs agrees qualitatively with the trends for PL intensities of surfactant-dispersed (6,5) SWNTs in water decorated with the same type of aryl defects.[18] In analogy to the 4-nitrophenyl (4-NO$_2$) case, the optimum reagent concentrations yielding the maximum PLQY were identified for the other substituents (**Supporting Information, Figure S8**). As expected from the reactivity pattern, only 0.037 mmol $L^{-1}$ of the electron-poor and very reactive 3,5-dichloro (3,5-Cl$_2$) reagent were required, as opposed to 3.7 mmol $L^{-1}$ of the electron-rich 4-methoxy (4-OMe) compound. The 4-bromo (4-Br) reagent has a medium reactivity and 0.37 mmol $L^{-1}$ were found to be optimal for maximum PL.

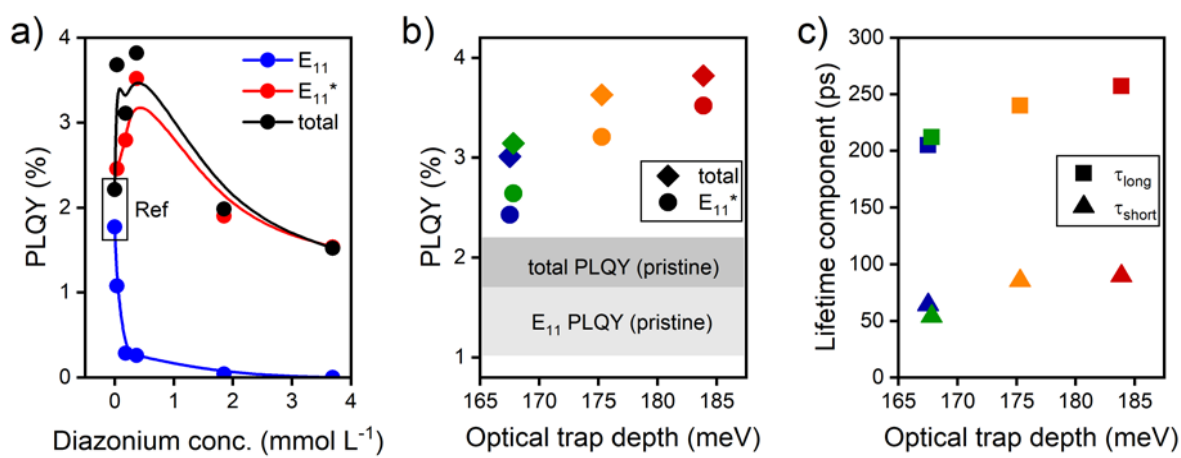

**Figure 4.** a) Spectral contributions to the PLQY *vs.* concentration of 4-nitrobenzenediazonium tetrafluoroborate as a measure of defect density. Solid lines are guides to the eye. b) Optimum PLQYs found for different substituents as a function of their optical trap depth. The corresponding substituents - in order of increasing optical trap depth - are 4-OMe (blue), 3,5-Cl$_2$ (green), 4-Br (orange) and 4-NO$_2$ (red). The gray shaded areas represent the typical ranges of E$_{11}$ and total PLQY of pristine (6,5) SWNTs dispersed by shear force mixing. c) Lifetime components ($\tau$) extracted from a biexponential fit to the time-resolved PL decay as a function of the optical trap depth. Note that PLQY and lifetime measurements were performed on the same samples.

In **Figure 4b** the maximum E$_{11}$* and total PLQYs are plotted as a function of the optical trap depth of the substituted aryl defect, *i.e.*, the difference between the E$_{11}$ and E$_{11}$* PL emission energies calculated from **Figures 2b** and **S4** and listed in **Table 1**. Even though the substituent on the aryl group modulates the optical trap depth by no more than ~10 %, it has a strong impact on the maximum PLQY. Evidently, both E$_{11}$* and total PLQY increase with the optical trap depth. Furthermore, the deeper the trap, the smaller the residual E$_{11}$ contribution (*i.e.*, difference between total and E$_{11}$* PLQY).

**Table 1.** Correlation of defect type (substituent on the aryl group), optical trap depth and emission properties.

| Defect type | Optical trap depth (meV) | Maximum total PLQY (%) | Long lifetime component (ps) |
|---|---|---|---|
| 4-OMe | 167.5 | 3.0 | 205 |
| 3,5-Cl$_2$ | 167.8 | 3.1 | 212 |
| 4-Br | 175.3 | 3.6 | 240 |
| 4-NO$_2$ | 183.9 | 3.8 | 257 |

This trend is best discussed in combination with an analysis of the PL decay dynamics. We recorded the PL decay at the respective peak wavelengths of the E$_{11}$* emission using time-correlated single-photon counting (TCSPC) and fitted a biexponential decay to the transients (refer to **Supporting Information, Figure S9 and S10** for a representative histogram and results for all substituents and defect densities). In agreement with literature,[26, 27] these $E_{11}$* PL decays feature a long-lived and a short-lived decay component ($\tau_{long}$ and $\tau_{short}$, respectively) with comparable amplitude. The short time constant has been interpreted as the time scale for redistribution of exciton population among bright and dark states localized at the defect.[26, 27] This redistribution period is followed-up by a slower decay of the trapped excitons through radiative and non-radiative channels.[26, 27] Consequently, the longer time constant is critical for the emission efficiency. Quantum chemical calculations suggest radiative lifetimes of ~2 ns for bright trap states,[29] but measured lifetimes are a few hundred picoseconds. Hence, recombination of trapped excitons is still dominated by non-radiative pathways. Specifically, localization-enhanced multi-phonon decay (MPD) and electronic-to-vibrational energy transfer (EVET) in the liquid phase were proposed as key mechanisms.[26, 27] The lifetime components found for the samples discussed in **Figure 4b** are shown in **Figure 4c** as a function of the optical trap depth. Note that as the PLQY is maximized at moderate defect densities, the lifetimes in this regime should also be the most reliable since there is neither a significant residual contribution from the $E_{11}$ phonon sideband, nor significant site-to-site interaction.

Similar to the trend in PLQY, $\tau_{long}$ is larger for deeper traps. Recently, it was observed that $\tau_{long}$ increases for a given SWNT chirality as longer-wavelength defect emission is probed.[26] It was concluded that thermal detrapping of excitons was responsible for this dependence because longer wavelengths are associated with deeper traps. In addition, the corresponding time scales on the order of ~100 ps can compete with MPD and EVET in most solvents.[26] In our case, the variation of the substituent on the aryl group causes a variation of the trap depth and thus, exciton lifetime. The values for $\tau_{long}$ range from 205 ps for 4-OMe- to 257 ps for 4-NO$_2$- substituted defects and agree well with those found for similar systems.[26, 27]

Concurrently to the long lifetime component, the total PLQY rises from 3.0 % to 3.8 % with the fraction of photons emitted *via* $E_{11}^*$ continuously increasing from 81 % to 92 %. If we assume that we compensated the differences in reactivity of the various diazonium salts by adjusting their concentrations, all samples compared in **Figure 4b** and **4c** should have similar defect densities. With this assumption we can attribute the substituent dependence of PLQY to the changes in trap depth, which affects the rate of thermal detrapping. The loss of defect state population suppresses the $E_{11}^*$ PLQY and the de-trapped mobile $E_{11}$ excitons are either quenched or decay radiatively, thereby increasing the $E_{11}$ PLQY, in accordance with the trend in **Figure 4b**. As the trap depth appears to be one of the factors limiting the PLQY in this study, the design of divalent dopants that could bind to the SWNT in an arrangement creating deeper traps is a promising route to even brighter SWNTs.[23, 36]

**Length Dependence of PL Brightening.** Apart from the parameters discussed so far - the defect density and trap depth - we find the PLQY to depend strongly on the quality of the initially dispersed SWNTs. To illustrate the variation of the PLQY in the starting material (shear-mixed, PFO-BPy wrapped (6,5) SWNTs), the typical range is indicated in **Figure 4b**. A similar degree of batch-to-batch variation is observed in *sp$^3$* functionalized (6,5) SWNTs with individual samples reaching total PLQYs up to 4.3 % and others less than 3 %. The general variation in SWNT dispersion quality is governed by many factors including the degree of mechanical damage of the SWNTs, coverage by dispersant, humidity and temperature. In the following, we discuss the role of tube length and initial exciton quenching site density on the PL brightening upon *sp$^3$* functionalization. We are particularly interested in this aspect with regard to the discrepancy between the two-fold PLQY enhancement observed here and previous reports of 5 to 10-fold brightening of (6,5) SWNTs that were dispersed in aqueous surfactant solution by tip sonication and sorted by gel chromatography.[18] Due to the harshness of tip sonication, the dispersed SWNTs are much shorter (average lengths of 350-700 nm)[26, 60] than shear-mixed SWNTs with average lengths of ~1.7 µm.[9]

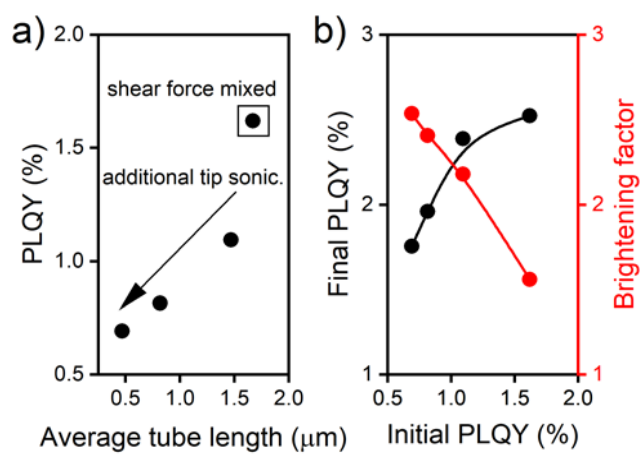

**Figure 5.** a) PLQY of PFO-BPy wrapped (6,5) SWNTs in toluene as a function of the average nanotube length. b) PLQY (black circles) after functionalization (final) with 4-bromo-benzenediazonium tetrafluoroborate *vs.* PLQY before functionalization (initial). The brightening factor (red circles) is defined as the ratio of final PLQY/initial PLQY. Solid lines are guides to the eye.

In addition to nanotube ends many types of sidewall defects also act as exciton quenching sites, thus promoting non-radiative relaxation and decreasing the PLQY depending on the dispersion method.[9, 14] To investigate the effect of higher quenching site densities (before any diazonium treatment) on the optical properties of the SWNTs after $sp^3$ functionalization without compromising the chiral purity of the sample, we shortened pre-sorted, shear-mixed (6,5) SWNTs by tip sonication. The degree of nanotube shortening was controlled by the duration of ultrasonication. The length distributions of the shear-mixed stock and tip-sonicated nanotube batches are provided in the **Supporting Information, Figure S11**. Atomic force microscopy statistics reveal that the average length drops from 1.7 µm in the shear-mixed stock down to 0.5 µm after the longest sonication. As shown in **Figure 5a**, the reduction in SWNT length leads to a significant decrease in PLQY from 1.6 % to 0.7 %. We find that short sonication times that result in marginal reductions of tube length still strongly affect the PLQY, which is most likely due to the creation of sidewall defects. After the shortening step, each batch was treated with the optimized concentration of 4-bromobenzenediazonium tetrafluoroborate (0.37 mmol L$^{-1}$) to produce (4-bromophenyl) $sp^3$ defects. Note that the shortening and functionalization steps were carried out in this order to account for reactivity differences in SWNTs with more or less damaged conjugated lattices. In this way, we mimic the scenario of starting materials dispersed under harsh conditions.

In **Figure 5b**, the spectrally integrated PLQY of $sp^3$ functionalized (6,5) SWNTs is plotted *versus* the initial PLQY of the corresponding batch of pristine polymer-wrapped SWNTs. It is evident that higher initial PLQYs also lead to higher post-functionalization PLQYs, which may be seen as the result of more efficient channeling of excitons to defect sites owing to a reduced probability of contact-quenching. However, the relative brightening, which we define as the ratio of final PLQY/initial PLQY, consistently follows an inverse trend. While the shortest SWNTs with low initial PLQY exhibit a ~2.5-fold brightening, the longest SWNTs are only brightened by a factor of ~1.5.

In summary, while high quality functionalized SWNTs feature the highest absolute PLQYs, the degree of brightening is higher for lower quality SWNTs. This observation is in agreement with the recent demonstration of photoluminescence from ultrashort and usually dark SWNTs by emissive trap sites.[34] As SWNTs with large numbers of non-radiative quenching sites will never display efficient $E_{11}$ emission due to fast exciton diffusion and contact-quenching,[14, 61] $sp^3$ defects allow a fraction of the exciton population to be harvested that otherwise would decay non-radiatively. The remaining discrepancy between our observed 2.5-fold brightening and the 7-fold brightening previously reported[18] for aqueous dispersions of (6,5) SWNTs with the same defect type likely stems from two factors. Firstly, the surfactant-dispersed SWNTs were even shorter on average[60] than the shortest batch studied in this work. Secondly, polymer-wrapped SWNTs usually have higher PLQYs than surfactant-stabilized SWNTs due to a higher degree of debundling and stronger dielectric shielding.[62] Untreated SWNTs dispersed in water suffer from faster non-radiative decay, which is alleviated by defect-induced exciton localization as this precludes diffusive sampling of the dielectric environment.[29] In polymer-wrapped SWNTs, however, the contrast between both cases should be smaller, thus functionalization results in a lower PLQY enhancement.

**PLQY and Brightening of SWNT Films.** If $sp^3$ functionalization is to be used to improve the quantum efficiency of SWNTs in light-emitting devices, it is crucial that the PLQY enhancement (relative to pristine SWNTs) is retained in a film with inter-tube and substrate interactions. The localization of excitons might even reduce the large PLQY drop from SWNT dispersions to thin films, which would be highly desirable. However, suitable deposition techniques for homogeneous films with defined thickness over large areas are required to reliably investigate and compare the thin film photoluminescence and PLQY of functionalized *vs.* pristine polymer-wrapped SWNTs. To achieve this, we scaled-up the functionalization process to reaction volumes of several hundred mL containing a total mass of ~0.5 mg of highly enriched, polymer-wrapped (6,5) SWNTs. Following our standard procedure, we tuned the defect density to the desired level corresponding to the maximum PLQY as determined above. Next, concentrated inks of functionalized and pristine SWNTs were prepared and used for aerosol-jet printing, which is an efficient method to deposit thick films in a controlled and spatially confined way.[63, 64] The low to moderate numbers of $sp^3$ defects introduced to the (6,5) SWNTs do not affect the colloidal stability or viscosity of the nanotube ink to an extent that would require adjustment of printing conditions. Hence, both pristine and functionalized (6,5) SWNTs were deposited under identical conditions. We printed 4 pairs of stripes (for comparison of pristine and functionalized SWNTs) with increasing thickness; all of them being visible with the bare eye. To date, there is no deposition technique for aqueous SWNT dispersions that enables the formation of comparably homogeneous, dense and large-area films, except filtration.[65] For a comparison of PL intensities, the respective stripes must be equally thick. Mapping of the Raman $G^+$ mode intensity confirmed that the stripes of each pair indeed have very similar thicknesses. With similar numbers of emitters, the PL intensity under identical excitation conditions can be used as a relative measure of PLQY. To achieve a direct comparison, the laser excitation spot was expanded to homogeneously illuminate a pair of printed stripes and the PL was imaged onto a 2D InGaAs camera. The resulting PL images of all printed stripes are shown in the **Supporting Information, Figure S12** together with the integrated Raman intensity and brightfield optical microscope images. The stripes of functionalized SWNTs are consistently brighter than the reference stripes. Careful analysis of the PL intensity as a function of film thickness shows that the films of functionalized SWNTs are on average a factor of 1.7 brighter than films of pristine SWNTs (see the **Supporting Information, Figure S13**). The PL micrograph of the thickest pair of stripes is shown in **Figure 6**. Clearly, the emission intensity of defect-functionalized SWNTs is higher. Subsequently, a grating was inserted to spectrally disperse the PL emission along one coordinate. The corresponding hyperspectral image is given next to the real-space image in **Figure 6**. The strong PL of the functionalized stripe originates from $E_{11}^*$ emission. Further, the asymmetry of the $E_{11}$ feature in contrast to the peak shape of the $E_{11}^*$ illustrates that apart from differences in the PLQY, the $E_{11}$ emission suffers from significant self-absorption due to the small Stokes shift, whereas the $E_{11}^*$ is nearly reabsorption-free.[60]

As a complementary test of enhanced emission efficiencies in films, we also spin-coated thin films of both samples with areas of ~1 cm$^2$ and measured the absolute PLQY in an integrating sphere in analogy to the liquid samples. Despite low emission intensities, the measurement was reproducible and PLQYs of 0.18 ± 0.05 % for the reference and 0.31 ± 0.06 % for the *sp$^3$* functionalized sample were found. The corresponding brightening factor of 1.7 matches the enhancement of PL intensity from the printed stripes exactly. Although the PL enhancement is not larger than in dispersion, it is at least retained for thin and thick functionalized SWNT films.

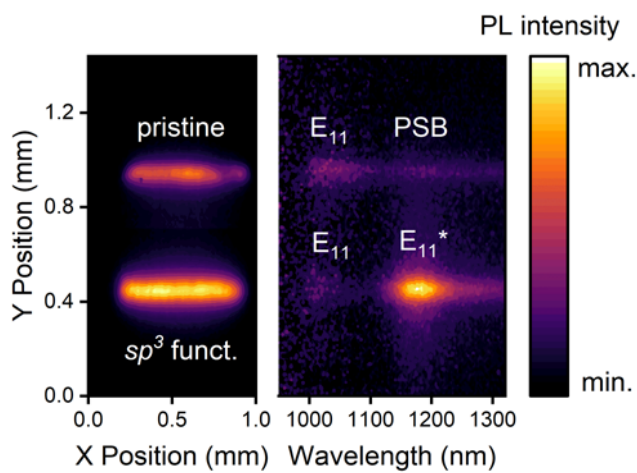

**Figure 6.** PL micrograph and hyperspectral image of printed stripes of pristine and 4-bromophenyl functionalized PFO-BPy-wrapped (6,5) SWNTs on glass recorded under continuous wave excitation at 640 nm.

**CONCLUSIONS**

We have presented a simple and scalable phase-transfer reaction scheme for the functionalization of polymer-wrapped SWNTs with luminescent $sp^3$ defects in organic nonhalogenated solvents based on aryldiazonium chemistry. As a result of the high monochiral purity and concentration of the functionalized nanotube dispersions, we were able to unambiguously identify the $E_{11}^*$ absorption of the $sp^3$ defects and found $E_{11}^*$ Stokes shifts of ~20 meV at low defect densities. We determined the absolute PLQY of the $sp^3$ functionalized (6,5) SWNTs in dispersion and probed its dependence on defect density and trap depth. Careful tuning of the defect density gave PLQYs of up to 4 % with ~90 % of photons emitted through the defect channel. The length and quality of the as-dispersed SWNTs strongly affected both the absolute PLQY and the relative brightening factor upon introduction of luminescent defects. While long SWNTs with high initial PLQY also gave the highest post-functionalization PLQY, the greatest relative brightening was found for the shortest nanotubes with the lowest initial PLQY. The presented method to functionalize polymer-wrapped SWNTs with $sp^3$ defects on a large scale in organic solvents further enabled the fabrication of homogeneous thin and thick nanotube films with bright and nearly reabsorption-free emission. While we only employed PFO-BPy-wrapped (6,5) nanotubes in this study, we presume that the general method can be applied likewise to other nanotube species with different wrapping polymers. The reproducible and large scale $sp^3$ functionalization of highly purified polymer-sorted semiconducting SWNTs will enable the fabrication of brighter carbon nanotube-based NIR emitting devices such as light-emitting diodes and field-effect transistors.

# METHODS

**Selective dispersion of (6,5) SWNTs.** As described previously,[9] (6,5) SWNTs were selectively extracted from CoMoCAT raw material (Chasm Advanced Materials, SG65i-L58, 0.38 g L$^{-1}$) by shear force mixing (Silverson L2/Air, 10,230 rpm, 72 h) and polymer-wrapping with poly-[(9,9-dioctylfluorenyl-2,7-diyl)-*alt-co*-(6,6')-(2,2'-bipyridine)] (PFO-BPy, American Dye Source, M$_w$ = 40 kg mol$^{-1}$, 0.5 g L$^{-1}$) in toluene. Aggregates were removed by centrifugation at 60,000 g (Beckman Coulter Avanti J26XP centrifuge) for 90 min and subsequent filtration (polytetrafluoroethylene (PTFE) syringe filter, 5 μm pore size).

**Shortening of (6,5) SWNTs.** Dispersions in toluene obtained from shear force mixing were tip-sonicated (Sonics, Vibracell VXC-500) using a tapered microtip at 35 % amplitude with 8 seconds on, 2 seconds off pulses at 5 °C for 4.5 to 23 hours to produce different degrees of nanotube shortening. The sonicated dispersions were centrifuged at 60,000 g for 45 min and the supernatant used for analysis.

***sp$^3$* Functionalization.** PFO-BPy-wrapped (6,5) SWNTs were covalently functionalized with a series of commercially available diazonium salts (4-bromo-, 4-methoxy, 4-nitro- and 3,5-dichlorobenzenediazonium tetrafluoroborate, Sigma Aldrich). For a detailed protocol, refer to the **Supporting Information**. Reactions were carried out at a (6,5) SWNT concentration of 0.36 mg L$^{-1}$ (corresponding to an E$_{11}$ absorbance of 0.2 for 1 cm path length)[66] in an 80:20 vol-% toluene/acetonitrile mixture. Briefly, a toluene solution of 18-crown-6 (18-crown-6, 99 %, Sigma Aldrich) was added to the as-prepared SWNT dispersion such that the 18-crown-6 concentration after addition of all other components was 7.6 mmol L$^{-1}$. The diazonium salt was dissolved in acetonitrile and an appropriate volume of this solution was added to the reaction vessel. After thorough mixing and a waiting period of 5 min, a solution of potassium acetate (KOAc) in 80:20 vol-% toluene/acetonitrile and 7.6 mmol L$^{-1}$ 18-crown-6 was added to the mixture to yield a KOAc concentration of ~$10^{-9}$ mol L$^{-1}$. The reaction proceeded at room temperature in the dark, stirring was not required. After typically 16 hours, the mixture was passed through a PTFE membrane filter (Merck Millipore, JVWP, 0.1 µm pore size) to collect the SWNTs. The filter cake was washed with acetonitrile and toluene to remove unreacted diazonium salt and excess polymer. Finally, the filter cake was redispersed in a small volume of pure toluene by bath sonication for 30 min.

**Printing of SWNTs.** Pristine and *sp*$^3$ functionalized (6,5) SWNTs were printed under identical conditions. Toluene and terpineol (mixture of isomers, Sigma Aldrich) were added to the dispersion to adjust the SWNT concentration to 5.4 mg L$^{-1}$ ($E_{11}$ absorbance of 3 cm$^{-1}$) and the terpineol concentration to 2 vol-%. This ink was used for aerosol jet printing (Aerosol Jet 200 printer, Optomec) with an ultrasonic atomizer.[63, 64] A 200 µm inner diameter nozzle was used at a sheath gas flow of 30 sccm and carrier gas flow of 25 sccm. The film thickness was tuned by the number of printing cycles. The sample stage was at 100 °C to facilitate fast evaporation of toluene. Residual terpineol was rinsed off with tetrahydrofuran and isopropanol.

**Optical measurements.** Absorption spectra were recorded with a Cary 6000i absorption spectrometer (Varian). Liquid phase PL spectra in the low excitation power density regime were collected with a HORIBA Jobin Yvon Fluorolog spectrofluorometer equipped with a 450 W xenon arc lamp and a liquid nitrogen cooled InGaAs line camera. For acquisition of PL spectra at higher excitation densities, a home-built setup was used. The sample was excited by the spectrally filtered output of a picosecond-pulsed supercontinuum laser source (Fianium WhiteLase SC400) focused by a 50× NIR-optimized objective (N.A. 0.65, Olympus). Scattered laser light was blocked by a dichroic long-pass filter (875 nm cut-off). The PL emission from the sample was dispersed by an Acton SpectraPro SP2358 spectrograph (grating 150 lines mm$^{-1}$) and detected with a liquid nitrogen cooled InGaAs line camera (Princeton Instruments

OMA V). For PL lifetime measurements using a time-correlated single photon counting scheme, the spectrally selected PL emission was focused onto a gated InGaAs/InP avalanche photodiode (Micro Photon Devices) *via* a 20× NIR-optimized objective (Mitutoyo). The PL quantum yield of dispersions and thin films was determined by an absolute method as reported earlier.[9] Samples were placed inside an integrating sphere (Labsphere) and the absorption of laser light at 575 nm ($E_{22}$ transition) as well as the PL emission were measured by transmitting the light through an optical fiber and coupling into the spectrometer. Photoluminescence images were acquired using an imaging spectrograph (Princeton Instruments IsoPlane SCT 320) and a thermoelectrically cooled 2D InGaAs camera (Princeton Instruments NIRvana). The sample was homogeneously illuminated with a 640 nm laser diode (Coherent OBIS, 40 mW continuous wave) and scattered laser light was blocked by a dichroic long-pass filter (850 nm cut-off).

**ACKNOWLEDGMENT**

This project has received funding from the European Research Council (ERC) under the European Union's Horizon 2020 research and innovation programme (Grant agreement No. 817494). J.L. acknowledges support by the Volkswagenstiftung (Grant No. 93404). V.L. thanks the DAAD-RISE programme. The authors thank Benjamin S. Flavel and Han Li for helpful discussions; and Arko Graf and Franziska Grün for initial input.

# REFERENCES

1.      Mueller, T.; Kinoshita, M.; Steiner, M.; Perebeinos, V.; Bol, A. A.; Farmer, D. B.; Avouris, P., Efficient Narrow-Band Light Emission from a Single Carbon Nanotube p-n Diode. *Nat. Nanotechnol.* **2009,** *5*, 27-31.

2.      Graf, A.; Murawski, C.; Zakharko, Y.; Zaumseil, J.; Gather, M. C., Infrared Organic Light-Emitting Diodes with Carbon Nanotube Emitters. *Adv. Mater.* **2018,** *30*, 1706711.

3.      Graf, A.; Held, M.; Zakharko, Y.; Tropf, L.; Gather, M. C.; Zaumseil, J., Electrical Pumping and Tuning of Exciton-Polaritons in Carbon Nanotube Microcavities. *Nat. Mater.* **2017,** *16*, 911-917.

4.      Jakubka, F.; Grimm, S. B.; Zakharko, Y.; Gannott, F.; Zaumseil, J., Trion Electroluminescence from Semiconducting Carbon Nanotubes. *ACS Nano* **2014,** *8*, 8477-8486.

5.      Avouris, P.; Freitag, M.; Perebeinos, V., Carbon-Nanotube Photonics and Optoelectronics. *Nat. Photon.* **2008,** *2*, 341-350.

6.      Wei, L.; Flavel, B. S.; Li, W.; Krupke, R.; Chen, Y., Exploring the Upper Limit of Single-Walled Carbon Nanotube Purity by Multiple-Cycle Aqueous Two-Phase Separation. *Nanoscale* **2017,** *9*, 11640-11646.

7.      O'Connell, M. J.; Bachilo, S. M.; Huffman, C. B.; Moore, V. C.; Strano, M. S.; Haroz, E. H.; Rialon, K. L.; Boul, P. J.; Noon, W. H.; Kittrell, C.; Ma, J.; Hauge, R. H.; Weisman, R. B.; Smalley, R. E., Band Gap Fluorescence from Individual Single-Walled Carbon Nanotubes. *Science* **2002,** *297*, 593-596.

8.      Samanta, S. K.; Fritsch, M.; Scherf, U.; Gomulya, W.; Bisri, S. Z.; Loi, M. A., Conjugated Polymer-Assisted Dispersion of Single-Wall Carbon Nanotubes: The Power of Polymer Wrapping. *Acc. Chem. Res.* **2014,** *47*, 2446-2456.

9.      Graf, A.; Zakharko, Y.; Schießl, S. P.; Backes, C.; Pfohl, M.; Flavel, B. S.; Zaumseil, J., Large Scale, Selective Dispersion of Long Single-Walled Carbon Nanotubes with High Photoluminescence Quantum Yield by Shear Force Mixing. *Carbon* **2016,** *105*, 593-599.

10.     Schneider, S.; Brohmann, M.; Lorenz, R.; Hofstetter, Y. J.; Rother, M.; Sauter, E.; Zharnikov, M.; Vaynzof, Y.; Himmel, H.-J.; Zaumseil, J., Efficient n-Doping and Hole Blocking in Single-Walled Carbon Nanotube Transistors with 1,2,4,5-Tetrakis(Tetramethylguanidino)Benzene. *ACS Nano* **2018,** *12*, 5895–5902.

11.     Rother, M.; Brohmann, M.; Yang, S.; Grimm, S. B.; Schießl, S. P.; Graf, A.; Zaumseil, J., Aerosol-Jet Printing of Polymer-Sorted (6,5) Carbon Nanotubes for Field-Effect Transistors with High Reproducibility. *Adv. Electron. Mater.* **2017,** *3*, 1700080.

12.     Bucella, S. G.; Salazar-Rios, J. M.; Derenskyi, V.; Fritsch, M.; Scherf, U.; Loi, M. A.; Caironi, M., Inkjet Printed Single-Walled Carbon Nanotube Based Ambipolar and Unipolar

Transistors for High-Performance Complementary Logic Circuits. *Adv. Electron. Mater.* **2016,** *2*, 1600094.

13. Mortimer, I. B.; Nicholas, R. J., Role of Bright and Dark Excitons in the Temperature-Dependent Photoluminescence of Carbon Nanotubes. *Phys. Rev. Lett.* **2007,** *98*, 027404.

14. Hertel, T.; Himmelein, S.; Ackermann, T.; Stich, D.; Crochet, J., Diffusion Limited Photoluminescence Quantum Yields in 1-D Semiconductors: Single-Wall Carbon Nanotubes. *ACS Nano* **2010,** *4*, 7161-7168.

15. Amori, A. R.; Hou, Z.; Krauss, T. D., Excitons in Single-Walled Carbon Nanotubes and Their Dynamics. *Annu. Rev. Phys. Chem.* **2018,** *69*, 81-99.

16. Ghosh, S.; Bachilo, S. M.; Simonette, R. A.; Beckingham, K. M.; Weisman, R. B., Oxygen Doping Modifies Near-Infrared Band Gaps in Fluorescent Single-Walled Carbon Nanotubes. *Science* **2010,** *330*, 1656-1660.

17. Miyauchi, Y.; Iwamura, M.; Mouri, S.; Kawazoe, T.; Ohtsu, M.; Matsuda, K., Brightening of Excitons in Carbon Nanotubes on Dimensionality Modification. *Nat. Photon.* **2013,** *7*, 715-719.

18. Piao, Y.; Meany, B.; Powell, L. R.; Valley, N.; Kwon, H.; Schatz, G. C.; Wang, Y., Brightening of Carbon Nanotube Photoluminescence through the Incorporation of $sp^3$ Defects. *Nat. Chem.* **2013,** *5*, 840-845.

19. Brozena, A. H.; Kim, M.; Powell, L. R.; Wang, Y., Controlling the Optical Properties of Carbon Nanotubes with Organic Colour-Centre Quantum Defects. *Nat. Rev. Chem.* **2019,** *3*, 375-392.

20. Luo, H. B.; Wang, P.; Wu, X.; Qu, H.; Ren, X.; Wang, Y., One-Pot, Large-Scale Synthesis of Organic Color Center-Tailored Semiconducting Carbon Nanotubes. *ACS Nano* **2019**, *13*, 8417-8424.

21. Kilina, S.; Ramirez, J.; Tretiak, S., Brightening of the Lowest Exciton in Carbon Nanotubes *via* Chemical Functionalization. *Nano. Lett.* **2012,** *12*, 2306-2312.

22. Gifford, B. J.; Kilina, S.; Htoon, H.; Doorn, S. K.; Tretiak, S., Exciton Localization and Optical Emission in Aryl-Functionalized Carbon Nanotubes. *J. Phys. Chem. C* **2018,** *122*, 1828-1838.

23. Shiraki, T.; Shiraishi, T.; Juhasz, G.; Nakashima, N., Emergence of New Red-Shifted Carbon Nanotube Photoluminescence Based on Proximal Doped-Site Design. *Sci. Rep.* **2016,** *6*, 28393.

24. Maeda, Y.; Minami, S.; Takehana, Y.; Dang, J. S.; Aota, S.; Matsuda, K.; Miyauchi, Y.; Yamada, M.; Suzuki, M.; Zhao, R. S.; Zhao, X.; Nagase, S., Tuning of the Photoluminescence and Up-Conversion Photoluminescence Properties of Single-Walled Carbon Nanotubes by Chemical Functionalization. *Nanoscale* **2016,** *8*, 16916-16921.

25.     Hartmann, N. F.; Yalcin, S. E.; Adamska, L.; Haroz, E. H.; Ma, X.; Tretiak, S.; Htoon, H.; Doorn, S. K., Photoluminescence Imaging of Solitary Dopant Sites in Covalently Doped Single-Wall Carbon Nanotubes. *Nanoscale* **2015,** *7*, 20521-20530.

26.     He, X.; Velizhanin, K. A.; Bullard, G.; Bai, Y.; Olivier, J. H.; Hartmann, N. F.; Gifford, B. J.; Kilina, S.; Tretiak, S.; Htoon, H.; Therien, M. J.; Doorn, S. K., Solvent- and Wavelength-Dependent Photoluminescence Relaxation Dynamics of Carbon Nanotube $sp^3$ Defect States. *ACS Nano* **2018,** *12*, 8060-8070.

27.     Hartmann, N. F.; Velizhanin, K. A.; Haroz, E. H.; Kim, M.; Ma, X.; Wang, Y.; Htoon, H.; Doorn, S. K., Photoluminescence Dynamics of Aryl $sp^3$ Defect States in Single-Walled Carbon Nanotubes. *ACS Nano* **2016,** *10*, 8355-8365.

28.     Kim, Y.; Velizhanin, K. A.; He, X.; Sarpkaya, I.; Yomogida, Y.; Tanaka, T.; Kataura, H.; Doorn, S. K.; Htoon, H., Photoluminescence Intensity Fluctuations and Temperature-Dependent Decay Dynamics of Individual Carbon Nanotube $sp^3$ Defects. *J. Phys. Chem. Lett.* **2019,** *10*, 1423-1430.

29.     He, X.; Sun, L.; Gifford, B. J.; Tretiak, S.; Piryatinski, A.; Li, X.; Htoon, H.; Doorn, S. K., Intrinsic Limits of Defect-State Photoluminescence Dynamics in Functionalized Carbon Nanotubes. *Nanoscale* **2019,** *11*, 9125-9132.

30.     Saha, A.; Gifford, B. J.; He, X.; Ao, G.; Zheng, M.; Kataura, H.; Htoon, H.; Kilina, S.; Tretiak, S.; Doorn, S. K., Narrow-Band Single-Photon Emission through Selective Aryl Functionalization of Zigzag Carbon Nanotubes. *Nat. Chem.* **2018,** *10*, 1089-1095.

31.     He, X.; Hartmann, N. F.; Ma, X.; Kim, Y.; Ihly, R.; Blackburn, J. L.; Gao, W.; Kono, J.; Yomogida, Y.; Hirano, A.; Tanaka, T.; Kataura, H.; Htoon, H.; Doorn, S. K., Tunable Room-Temperature Single-Photon Emission at Telecom Wavelengths from $sp^3$ Defects in Carbon Nanotubes. *Nat. Photon.* **2017,** *11*, 577-582.

32.     He, X.; Gifford, B. J.; Hartmann, N. F.; Ihly, R.; Ma, X.; Kilina, S. V.; Luo, Y.; Shayan, K.; Strauf, S.; Blackburn, J. L.; Tretiak, S.; Doorn, S. K.; Htoon, H., Low-Temperature Single Carbon Nanotube Spectroscopy of $sp^3$ Quantum Defects. *ACS Nano* **2017,** *11*, 10785-10796.

33.     He, X.; Htoon, H.; Doorn, S. K.; Pernice, W. H. P.; Pyatkov, F.; Krupke, R.; Jeantet, A.; Chassagneux, Y.; Voisin, C., Carbon Nanotubes as Emerging Quantum-Light Sources. *Nat. Mater.* **2018,** *17*, 663-670.

34.     Danné, N.; Kim, M.; Godin, A. G.; Kwon, H.; Gao, Z.; Wu, X.; Hartmann, N. F.; Doorn, S. K.; Lounis, B.; Wang, Y.; Cognet, L., Ultrashort Carbon Nanotubes That Fluoresce Brightly in the Near-Infrared. *ACS Nano* **2018,** *12*, 6059-6065.

35.     Kwon, H.; Kim, M.; Meany, B.; Piao, Y.; Powell, L. R.; Wang, Y., Optical Probing of Local pH and Temperature in Complex Fluids with Covalently Functionalized, Semiconducting Carbon Nanotubes. *J. Phys. Chem. C* **2015,** *119*, 3733-3739.

36. Kwon, H.; Furmanchuk, A.; Kim, M.; Meany, B.; Guo, Y.; Schatz, G. C.; Wang, Y., Molecularly Tunable Fluorescent Quantum Defects. *J. Am. Chem. Soc.* **2016,** *138*, 6878-6885.

37. Li, H.; Gordeev, G.; Garrity, O.; Reich, S.; Flavel, B. S., Separation of Small-Diameter Single-Walled Carbon Nanotubes in One to Three Steps with Aqueous Two-Phase Extraction. *ACS Nano* **2019,** *13*, 2567-2578.

38. Cui, J.; Su, W.; Yang, D.; Li, S.; Wei, X.; Zhou, N.; Zhou, W.; Xie, S.; Kataura, H.; Liu, H., Mass Production of High-Purity Semiconducting Carbon Nanotubes by Hydrochloric Acid Assisted Gel Chromatography. *ACS Appl. Nano Mater.* **2018,** *2*, 343-350.

39. Derenskyi, V.; Gomulya, W.; Gao, J.; Bisri, S. Z.; Pasini, M.; Loo, Y.-L.; Loi, M. A., Semiconducting SWNTs Sorted by Polymer Wrapping: How Pure Are They? *Appl. Phys. Lett.* **2018,** *112*.

40. Hennrich, F.; Li, W.; Fischer, R.; Lebedkin, S.; Krupke, R.; Kappes, M. M., Length-Sorted, Large-Diameter, Polyfluorene-Wrapped Semiconducting Single-Walled Carbon Nanotubes for High-Density, Short-Channel Transistors. *ACS Nano* **2016,** *10*, 1888-1895.

41. Bahr, J. L.; Tour, J. M., Highly Functionalized Carbon Nanotubes Using *in situ* Generated Diazonium Compounds. *Chem. Mater.* **2001,** *13*, 3823-3824.

42. Wang, J.; Shea, M. J.; Flach, J. T.; McDonough, T. J.; Way, A. J.; Zanni, M. T.; Arnold, M. S., Role of Defects as Exciton Quenching Sites in Carbon Nanotube Photovoltaics. *J. Phys. Chem. C* **2017,** *121*, 8310-8318.

43. Beadle, J. R.; Korzeniowski, S. H.; Rosenberg, D. E.; Garcia-Slanga, B. J.; Gokel, G. W., Phase-Transfer-Catalyzed Gomberg-Bachmann Synthesis of Unsymmetrical Biarenes: A Survey of Catalysts and Substrates. *J. Org. Chem.* **1984,** *49*, 1594-1603.

44. Usrey, M. L.; Lippmann, E. S.; Strano, M. S., Evidence for a Two-Step Mechanism in Electronically Selective Single-Walled Carbon Nanotube Reactions. *J. Am. Chem. Soc.* **2005,** *127*, 16129-16135.

45. Shiraki, T.; Niidome, Y.; Toshimitsu, F.; Shiraishi, T.; Shiga, T.; Yu, B.; Fujigaya, T., Solvatochromism of near Infrared Photoluminescence from Doped Sites of Locally Functionalized Single-Walled Carbon Nanotubes. *Chem. Commun.* **2019,** *55*, 3662-3665.

46. Jones, M.; Engtrakul, C.; Metzger, W. K.; Ellingson, R. J.; Nozik, A. J.; Heben, M. J.; Rumbles, G., Analysis of Photoluminescence from Solubilized Single-Walled Carbon Nanotubes. *Phys. Rev. B* **2005,** *71*, 115426.

47. Kim, M.; Adamska, L.; Hartmann, N. F.; Kwon, H.; Liu, J.; Velizhanin, K. A.; Piao, Y.; Powell, L. R.; Meany, B.; Doorn, S. K.; Tretiak, S.; Wang, Y., Fluorescent Carbon Nanotube Defects Manifest Substantial Vibrational Reorganization. *J. Phys. Chem. C* **2016,** *120*, 11268-11276.

48. Ma, X.; Adamska, L.; Yamaguchi, H.; Yalcin, S. E.; Tretiak, S.; Doorn, S. K.; Htoon, H., Electronic Structure and Chemical Nature of Oxygen Dopant States in Carbon Nanotubes. *ACS Nano* **2014,** *8*, 10782-10789.

49. Gifford, B. J.; Sifain, A. E.; Htoon, H.; Doorn, S. K.; Kilina, S.; Tretiak, S., Correction Scheme for Comparison of Computed and Experimental Optical Transition Energies in Functionalized Single-Walled Carbon Nanotubes. *J. Phys. Chem. Lett.* **2018,** *9*, 2460-2468.

50. Dresselhaus, M. S.; Jorio, A.; Souza Filho, A. G.; Saito, R., Defect Characterization in Graphene and Carbon Nanotubes Using Raman Spectroscopy. *Phil. Trans. R. Soc. A* **2010,** *368*, 5355-5377.

51. Pfohl, M.; Tune, D. D.; Graf, A.; Zaumseil, J.; Krupke, R.; Flavel, B. S., Fitting Single-Walled Carbon Nanotube Optical Spectra. *ACS Omega* **2017,** *2*, 1163-1171.

52. Naumov, A. V.; Ghosh, S.; Tsyboulski, D. A.; Bachilo, S. M.; Weisman, R. B., Analyzing Absorption Backgrounds in Single-Walled Carbon Nanotube Spectra. *ACS Nano* **2011,** *5*, 1639-1648.

53. Nair, N.; Usrey, M. L.; Kim, W.-J.; Braatz, R. D.; Strano, M. S., Estimation of the (n,m) Concentration Distribution of Single-Walled Carbon Nanotubes from Photoabsorption Spectra. *Anal. Chem.* **2006,** *78*, 7689-7696.

54. Wang, F.; Dukovic, G.; Knoesel, E.; Brus, L. E.; Heinz, T. F., Observation of Rapid Auger Recombination in Optically Excited Semiconducting Carbon Nanotubes. *Phys. Rev. B* **2004,** *70*, 241403.

55. Wang, F.; Wu, Y.; Hybertsen, M. S.; Heinz, T. F., Auger Recombination of Excitons in One-Dimensional Systems. *Phys. Rev. B* **2006,** *73*, 245424.

56. Iwamura, M.; Akizuki, N.; Miyauchi, Y.; Mouri, S.; Shaver, J.; Gao, Z.; Cognet, L.; Lounis, B.; Matsuda, K., Nonlinear Photoluminescence Spectroscopy of Carbon Nanotubes with Localized Exciton States. *ACS Nano* **2014,** *8*, 11254-11260.

57. Hartleb, H.; Späth, F.; Hertel, T., Evidence for Strong Electronic Correlations in the Spectra of Gate-Doped Single-Wall Carbon Nanotubes. *ACS Nano* **2015,** *9*, 10461-10470.

58. Stürzl, N.; Lebedkin, S.; Kappes, M. M., Revisiting the Laser Dye Styryl-13 as a Reference Near-Infrared Fluorophore: Implications for the Photoluminescence Quantum Yields of Semiconducting Single-Walled Carbon Nanotubes. *J. Phys. Chem. A* **2009,** *113*, 10238-10240.

59. Doyle, C. D.; Rocha, J.-D. R.; Weisman, R. B.; Tour, J. M., Structure-Dependent Reactivity of Semiconducting Single-Walled Carbon Nanotubes with Benzenediazonium Salts. *J. Am. Chem. Soc.* **2008,** *130*, 6795-6800.

60. Powell, L. R.; Piao, Y.; Ng, A. L.; Wang, Y., Channeling Excitons to Emissive Defect Sites in Carbon Nanotube Semiconductors Beyond the Dilute Regime. *J. Phys. Chem. Lett.* **2018,** *9*, 2803-2807.

61. Harrah, D. M.; Swan, A. K., The Role of Length and Defects on Optical Quantum Efficiency and Exciton Decay Dynamics in Single-Walled Carbon Nanotubes. *ACS Nano* **2011,** *5*, 647-655.

62. Nish, A.; Hwang, J. Y.; Doig, J.; Nicholas, R. J., Highly Selective Dispersion of Single-Walled Carbon Nanotubes Using Aromatic Polymers. *Nat. Nanotechnol.* **2007,** *2*, 640-646.

63. Berger, F. J.; Higgins, T. M.; Rother, M.; Graf, A.; Zakharko, Y.; Allard, S.; Matthiesen, M.; Gotthardt, J. M.; Scherf, U.; Zaumseil, J., From Broadband to Electrochromic Notch Filters with Printed Monochiral Carbon Nanotubes. *ACS Appl. Mater. Interfaces* **2018,** *10*, 11135-11142.

64. Rother, M.; Brohmann, M.; Yang, S.; Grimm, S. B.; Schießl, S. P.; Graf, A.; Zaumseil, J., Aerosol-Jet Printing of Polymer-Sorted (6,5) Carbon Nanotubes for Field-Effect Transistors with High Reproducibility. *Adv. Electron. Mater.* **2017**, 1700080.

65. He, X.; Gao, W.; Xie, L.; Li, B.; Zhang, Q.; Lei, S.; Robinson, J. M.; Hároz, E. H.; Doorn, S. K.; Wang, W.; Vajtai, R.; Ajayan, P. M.; Adams, W. W.; Hauge, R. H.; Kono, J., Wafer-Scale Monodomain Films of Spontaneously Aligned Single-Walled Carbon Nanotubes. *Nat. Nanotechnol.* **2016,** *11*, 633–638.

66. Streit, J. K.; Bachilo, S. M.; Ghosh, S.; Lin, C. W.; Weisman, R. B., Directly Measured Optical Absorption Cross Sections for Structure-Selected Single-Walled Carbon Nanotubes. *Nano Lett.* **2014,** *14*, 1530-1536.

# Supporting Information

## Table of Contents

## Dispersion Stability in Reaction Medium

Absorption and PL spectra confirm that there is no substantial aggregation of (6,5) SWNTs/PFO-BPy in the solvent mixture (toluene/acetonitrile) used for the reaction. Most importantly, the absorption spectrum does not show a significant increase in the scattering background. The $E_{11}$ transition is only marginally broadened in absorption and emission.

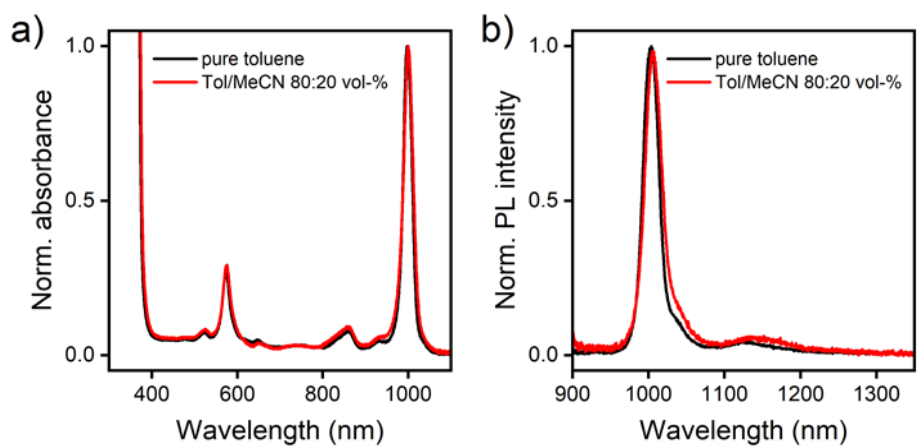

**Figure S1.** Normalized a) absorption and b) PL spectra of PFO-BPy-wrapped (6,5) SWNTs at a concentration of 0.36 mg L$^{-1}$ (corresponding to an $E_{11}$ absorbance of 0.2 in a 1 cm cuvette) in pure toluene and the solvent mixture used for the reaction (toluene/acetonitrile 80:20 vol%).

**Mechanistic Considerations and Impact of KOAc on Functionalization**

The reaction between aryldiazonium salts and SWNTs proceeds in two steps: (1) the generation of aryl radicals and (2) the addition of these radicals to the conjugated SWNT lattice.

In water, as long as the solution is not too acidic (pH > 5), the aryl radicals are produced *via* the Gomberg-Bachmann mechanism.[1] This process involves the reaction with hydroxide ions, diazonium anhydride formation and its decomposition to aryl radicals.[2] In acidic solutions, this pathway is suppressed and aryl radicals are instead created *via* a single-electron transfer from the SWNT to the diazonium salt, followed by the release of $N_2$.[2, 3] In this case, the reaction proceeds more slowly than upon Gomberg-Bachmann initiation.[2]

In organic solvents, the classical Gomberg-Bachmann pathway involving hydroxide ions is obviously unavailable. However, Beadle *et al.* reported that carboxylate salts, such as potassium acetate, can take up the role of hydroxide ions in the Gomberg-Bachmann mechanism and serve as efficient aryl radical initiators in organic solvents.[4]

This prompted us to explore the effect of potassium acetate (KOAc) on *sp³* functionalization in our system. Note that even in the presence of an ether crown the solubility of KOAc in organic solvents remains very low so that only small concentrations in the reaction mixture are attainable. Upon addition of KOAc to the mixture of (6,5) SWNTs and diazonium salt, the solution immediately turned yellow (at sufficiently high KOAc and diazonium salt concentrations) suggesting fast generation of aryl radicals and their homocoupling to larger conjugated systems that absorb light in the visible spectral range. PL spectra of (6,5) SWNTs functionalized with 4-methoxybenzenediazonium tetrafluoroborate in the absence and in the presence of low and high concentrations of KOAc are compared in **Figure S2**. Surprisingly, we observe very similar densities of luminescent defects in all cases, as estimated from the $E_{11}*$ to $E_{11}$ PL signal ratio. However, the sample functionalized at low KOAc concentration shows a slightly smaller contribution from deeper exciton traps ($E_{11}*^-$), which are unwanted here because they broaden the main emission signal. Hence, we performed all other experiments in this work with a low concentration ($10^{-9}$ mol L$^{-1}$) of KOAc. For comparison, *sp³* functionalization in aqueous dispersions is typically carried out at pH~5.5,[5] thus the concentration of initiating hydroxide ions is ~$10^{-8.5}$ mol L$^{-1}$. However, the case of aqueous dispersions is different since hydroxide ions can be replenished by autoprotolysis of water. On the other hand, small amounts of acetate may be regenerated by hydrolysis of the product acetic anhydride[4] depending on the humidity level. The fact that the addition of KOAc does not increase the efficiency of defect creation, but leads to changes in selectivity, might indicate that other activation pathways, such as electron-transfer from the SWNT to the diazonium salt,[2, 3] dominate in this solvent system.

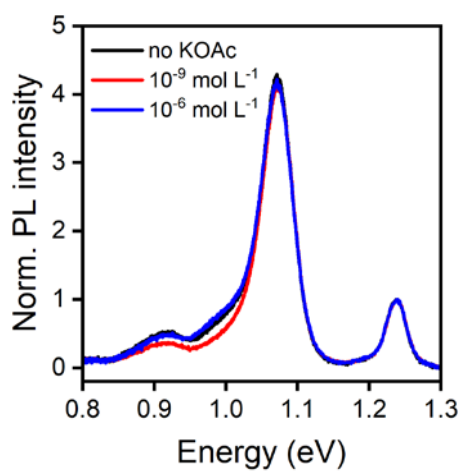

**Figure S2.** PL spectra of toluene dispersions of (6,5) SWNTs functionalized with 4-methoxybenzenediazonium tetrafluoroborate in the absence of KOAc (black curve), at a low KOAc concentration (~$10^{-9}$ mol L$^{-1}$, red curve) and at a high concentration of KOAc (~$10^{-6}$ mol L$^{-1}$, blue curve). Measurements were performed under pulsed excitation (575 nm, ~0.5 mJ cm$^{-2}$). Note that the spectra are displayed on a linear energy scale to allow for better comparison of peak widths.

## Detailed *sp³* Functionalization Protocol

**Method.** PFO-BPy-wrapped (6,5) SWNTs are *sp³* functionalized in a toluene/acetonitrile 80:20 vol-% mixture by reaction with aryldiazonium salts in the presence of 18-crown-6 as a phase-transfer agent. The density of created *sp³* defects is tuned by varying the concentration of aryldiazonium salt in the reaction mixture. A low concentration of potassium acetate can be added to increase the selectivity.

**Starting material.** Dispersions of PFO-BPy wrapped (6,5) SWNTs in toluene are the starting material for this functionalization procedure. The (6,5) SWNT concentration in the reaction mixture was always 0.36 mg L$^{-1}$ (corresponding to an E$_{11}$ absorbance of 0.2 for a 1 cm cuvette).[6] Free PFO-BPy aggregates when the acetonitrile content in the reaction mixture is too high, hence, it is important to consider the PFO-BPy concentration in the starting dispersion of (6,5) SWNTs. The typical PFO-BPy concentrations used for SWNT dispersion *via* polymer-wrapping are 0.5 to 2 g L$^{-1}$. For our shear force mixing approach, with which all starting dispersions were prepared, we use 0.5 g L$^{-1}$. Since the concentration of dispersed (6,5) SWNTs after mixing is typically around 1.8 mg L$^{-1}$ (corresponding to an E$_{11}$ absorbance of 1 cm$^{-1}$), the (6,5) SWNTs and free PFO-BPy are usually ~5-fold diluted, leading to an estimated PFO-BPy concentration in the reaction mixture of ~0.1 g L$^{-1}$. We do not observe significant aggregation at this PFO-BPy concentration in a toluene/acetonitrile 80:20 vol-% mixture (compare Figure S1). If a higher PFO-BPy concentration is used in the initial dispersion step, a higher dilution of (6,5) SWNTs and free PFO-BPy (or removal of excess PFO-BPy by washing) may be necessary. If the concentration of (6,5) SWNTs in the reaction mixture is reduced, the concentration of aryldiazonium salt must be reduced by the same factor in order to achieve a similar defect density as in this report.

**Reagents.** All chemicals were purchased from Sigma Aldrich and used without further purification. Specifically, 4-bromobenzenediazonium tetrafluoroborate (96 %) 4-methoxy-benzenediazonium tetrafluoroborate (98 %), 4-nitrobenzenediazonium tetrafluoroborate (97 %) and 3,5-dichlorophenyldiazonium tetrafluoroborate, potassium acetate (98 %) and 18-crown-6 (99 %).

**Phase-transfer approach.** 18-crown-6 strongly improves the solubility of diazonium salts in organic solvents. In many phase-transfer reactions, catalytic amounts of ether crown are sufficient to allow the reaction to proceed. However, due to the low efficiency of the reaction between diazonium salts and SWNTs (in water ~10 % and substantially lower in organic media), molecules of ether crown are permanently occupied by unreacted diazonium cations, thereby prohibiting a catalytic cycle. Hence, the molar concentration of ether crown should be equal to or higher than that of the diazonium salt. For best comparability, the concentration of 18-crown-6 in the reaction was always 7.6 mmol $L^{-1}$ (2 g $L^{-1}$) in this work.

**Stepwise protocol**

1) Prepare a solution of 18-crown-6 in toluene such that the 18-crown-6 concentration after addition of all components, *i.e.* in the final reaction volume, becomes 7.6 mmol $L^{-1}$. Add this solution to an appropriate volume of (6,5) SWNT dispersion in toluene.

2) Add pure acetonitrile to the mixture such that the mixing ratio of toluene/acetonitrile becomes 80:20 vol-% after addition of the remaining components.

3) Prepare a stock solution of diazonium salt in pure acetonitrile at room temperature in an amber vial (to avoid light-induced reactions). Add an appropriate volume of this diazonium salt solution to the reaction vessel and mix thoroughly. Note that all diazonium salt concentrations stated in this work apply to the reaction mixture and not to the stock solution. If the reaction is to be conducted without potassium acetate (KOAc), the mixture should be protected from light and allowed to react for usually ~16 hours (stirring is not required). Then, continue with step 5).

4) If the reaction is to be conducted in the presence of KOAc, a stock solution must be prepared beforehand. The stock solution is prepared by dissolving KOAc (0.01 g L$^{-1}$, ~10$^{-4}$ mol L$^{-1}$) in a solution of 18-crown-6 (7.6 mmol L$^{-1}$) in toluene/acetonitrile 80:20 vol-%. Due to the poor solubility, KOAc is directly dissolved in the final solvent system to avoid uncontrolled precipitation. The stock solution is usually further diluted and an appropriate volume is added to the reaction mixture to produce a KOAc concentration of ~10$^{-9}$ mol L$^{-1}$. After addition of the diazonium salt and thorough mixing, one should wait for ~5 min before adding the KOAc solution. This is to allow the diazonium salt to adsorb on the SWNT surface before adding the KOAc as a radical initiator. Then, the reaction vessel is protected from light and the reaction allowed to proceed for usually ~16 hours (stirring is not required).

5) The reaction mixture is passed through a PTFE membrane filter (*e.g.* Merck Millipore, JVWP, 0.1 μm pore size) to collect the SWNTs and the filter cake is washed with acetonitrile (~10 mL) and toluene (~5 mL) to remove unreacted diazonium salt and excess polymer. The washing step is crucial to recover photoluminescence of the SWNTs.

6) Finally, the filter cake is redispersed in a small volume of pure toluene by bath sonication for 30 min.

**E$_{11}$\* Absorption and Emission Maxima and Stokes Shifts**

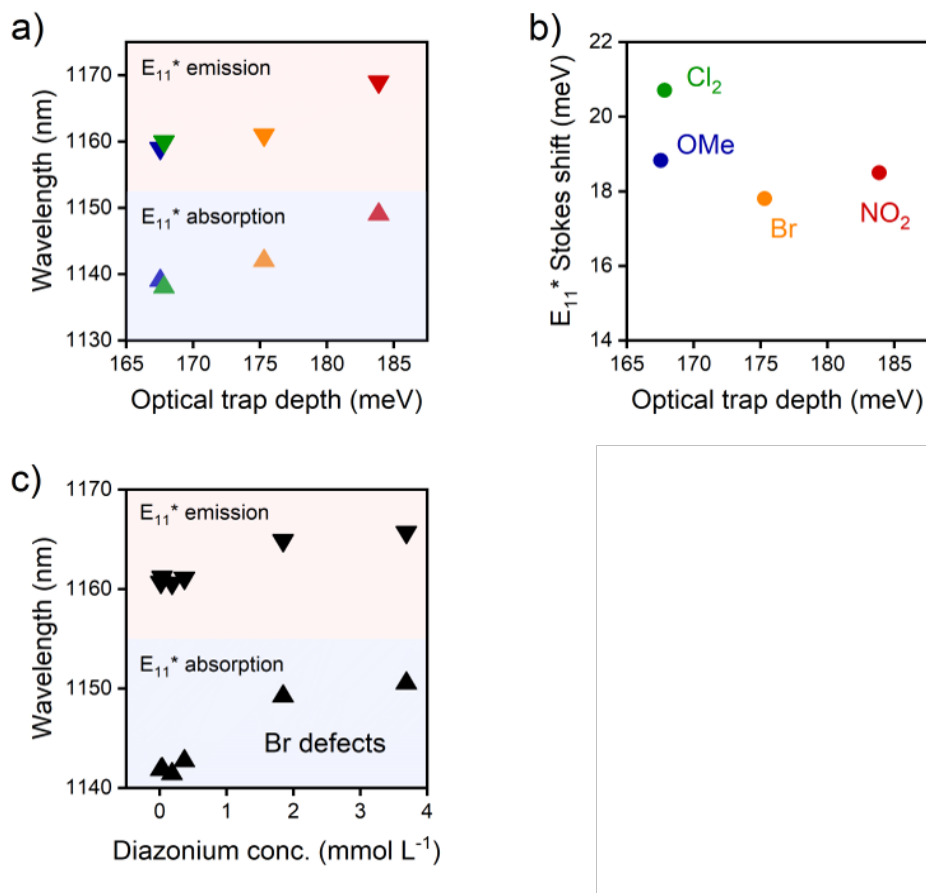

**Figure S3.** a) E$_{11}$\* maximum absorption and emission wavelengths for (6,5) SWNTs functionalized with different types of aryl defects (4-OMe; 3,5-Cl$_2$; 4-Br; 4-NO$_2$). All data points correspond to low defect densities. b) Corresponding E$_{11}$\* Stokes shifts for the samples in a). c) E$_{11}$\* maximum absorption and emission wavelengths for (6,5) SWNTs functionalized with different densities of 4-Br-phenyl defects given by the diazonium salt concentration in the reaction volume.

# PL Spectra of (6,5) SWNTs Functionalized with Different Aryl Defects

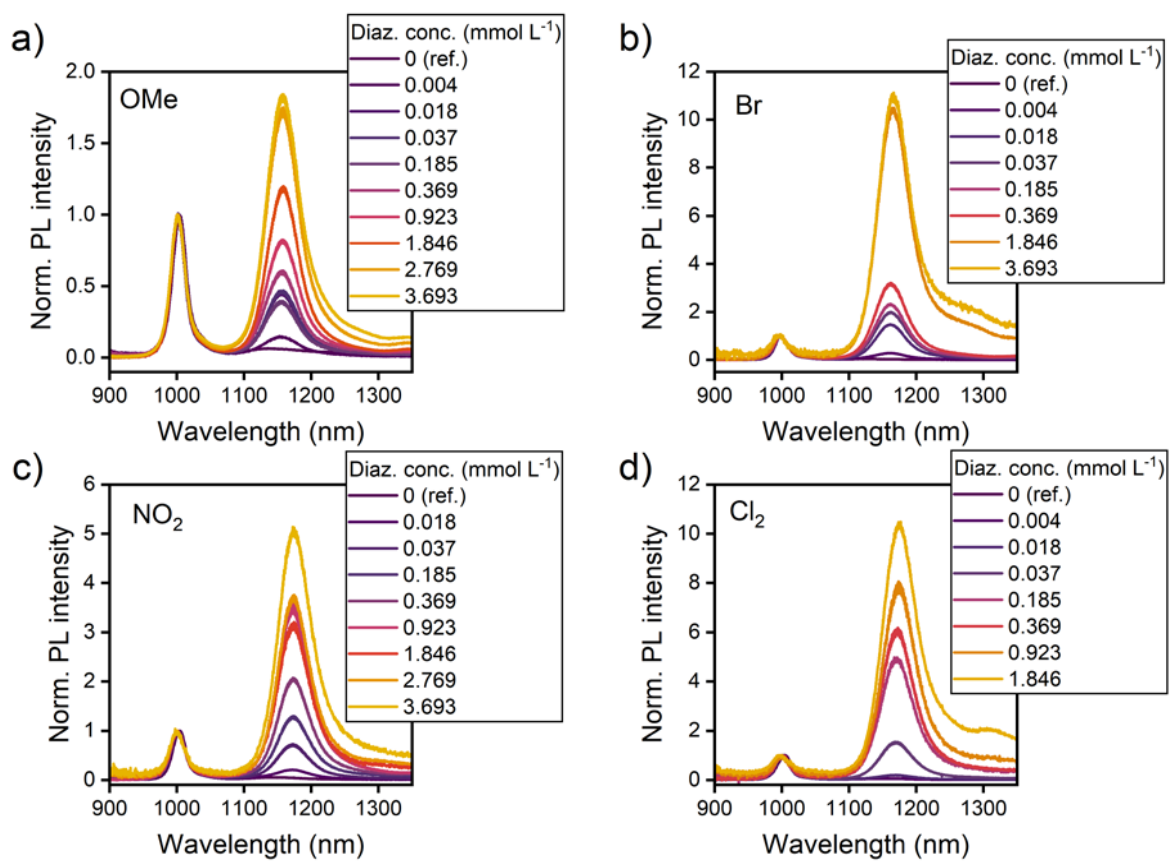

**Figure S4.** PL spectra of dispersions of (6,5) SWNTs functionalized with different types of aryl defects (4-OMe; 3,5-$Cl_2$; 4-Br; 4-$NO_2$) at various defect densities. Measurements were performed under pulsed excitation (575 nm, ~0.5 mJ cm$^{-2}$), spectra were normalized to the $E_{11}$ emission.

# Raman Spectra of Functionalized (6,5) SWNTs

(6,5) SWNTs were drop-cast on glass slides and measured with a Renishaw inVia Reflex confocal Raman microscope equipped with a 50× long working distance objective (N.A. 0.5, Olympus). The films were mapped by scanning with a 532 nm laser. The average of ~1500 spectra for each sample is presented in **Figure S5**. The slight asymmetry of the D mode in **Figure S5b** is likely due to bundling of SWNTs during drop-casting.[7]

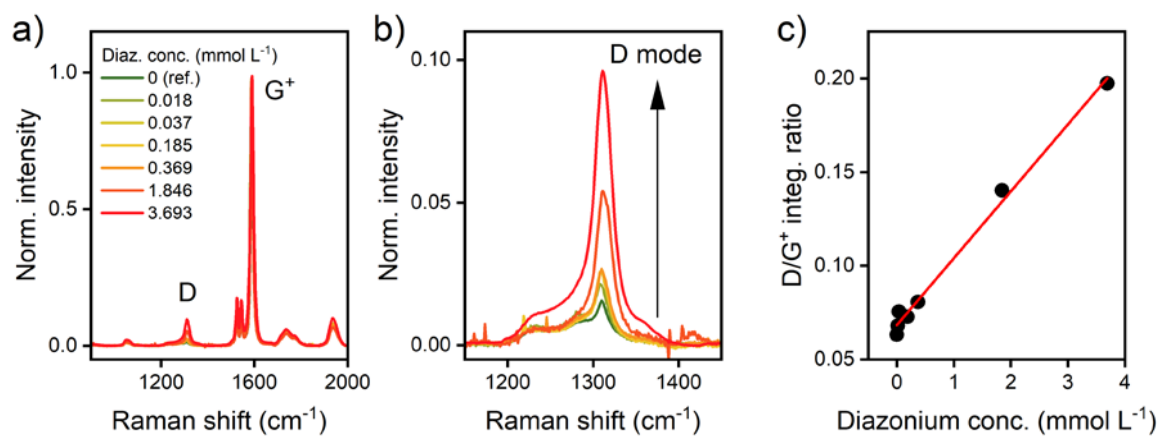

**Figure S5.** a) Averaged Raman spectra of drop-cast films of (6,5) SWNTs functionalized with different densities of 4-bromophenyl defects. b) Zoom-in on the D mode region. c) $D/G^+$ area ratio as a function of diazonium salt concentration in the reaction mixture and linear fit to the data.

## Measures of Defect Densities

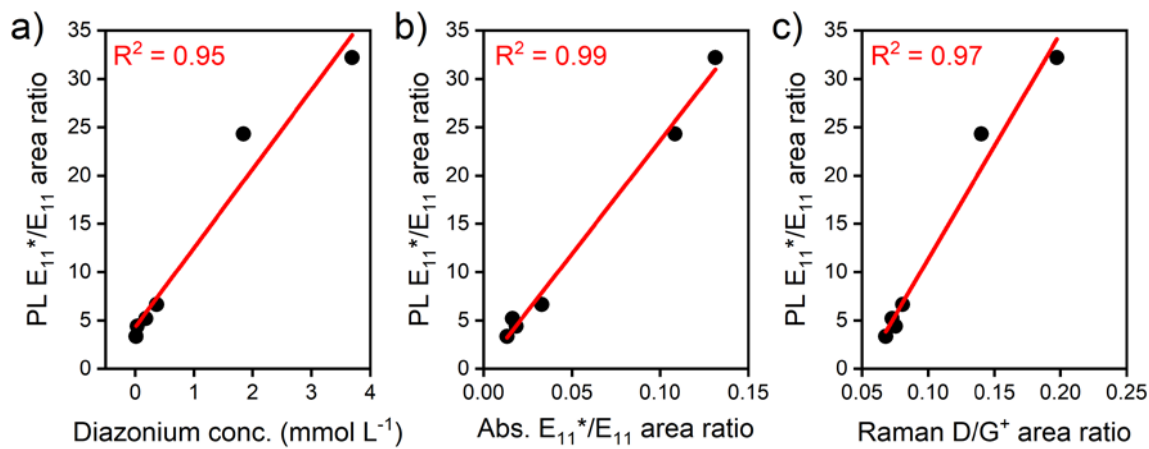

**Figure S6.** Absorption, PL and Raman properties of functionalized (6,5) SWNTs using different concentrations of 4-bromobenzenediazonium tetrafluoroborate. Before integrating the $E_{11}$ and $E_{11}^*$ absorption bands, a scattering background of the form $S(\lambda) = S_0 \exp(-b\lambda)$ was fitted to the absorption spectrum and subtracted.[8, 9] a) Ratio of integrated PL signals $E_{11}^*/E_{11}$ vs. diazonium salt concentration and linear fit to the data. b) Ratio of integrated PL signals $E_{11}^*/E_{11}$ vs. ratio of integrated absorbance $E_{11}^*/E_{11}$ and linear fit to the data. c) Ratio of integrated PL signals $E_{11}^*/E_{11}$ vs. Raman D/G$^+$ ratio and linear fit to the data. $R^2$ represents the coefficient of determination for the linear fits.

**Power-Dependence of Photoluminescence**

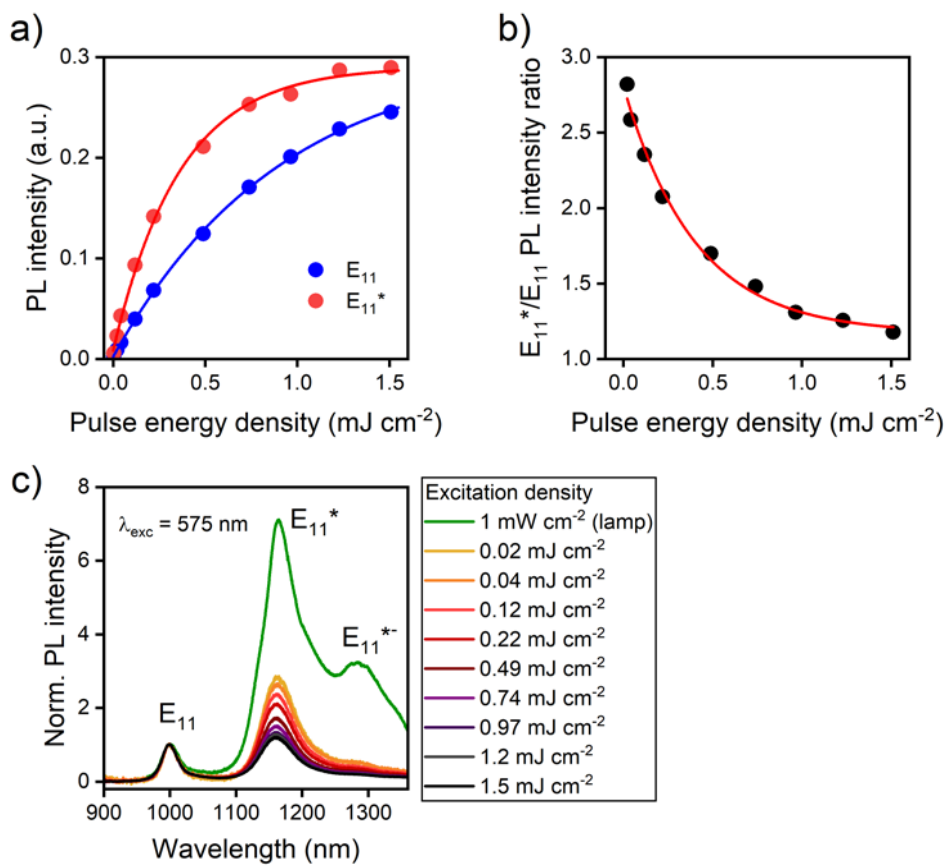

**Figure S7.** Power-dependent photoluminescence data for (6,5) SWNTs functionalized with 4-bromophenyl defects. a) Intensity of $E_{11}$ and $E_{11}*$ emission *vs.* pulse energy per area and monoexponential fits to the data. b) $E_{11}*/E_{11}$ PL peak intensity ratio *vs.* pulse energy per area and monoexponential fit to the data. c) PL spectra recorded at different excitation densities, including lamp illumination and pulsed laser excitation (~6 ps pulse width). The excitation wavelength was 575 nm in all cases.

# Photoluminescence Quantum Yield Measurements

**Measurement procedure.** The PL quantum yield (PLQY) of dispersions and thin films was determined by an absolute method as reported earlier.[10] When liquid samples were measured, 1 mL of dispersion was filled into a quartz cuvette (Hellma Analytics, QX type). Samples were placed inside an integrating sphere (Labsphere) and excited at 575 nm ($E_{22}$ transition) by the spectrally-filtered output of a pulsed supercontinuum laser source (Fianium WhiteLase SC400, 20 MHz repetition rate, ~6 ps pulse width). In this configuration, the absorption of laser light and the photoluminescence were recorded by transmitting the light through an optical fiber and coupling into the spectrometer. To account for absorption of laser light by the solvent (toluene) and the cuvette, the same measurement was performed on a cuvette filled with 1 mL of toluene. Integration of the laser signal intensity relative to that of the toluene reference yields a value proportional to the number of absorbed photons. Analogously, integration of the PL spectrum yields a value proportional to the number of emitted photons. Finally, the PLQY is calculated as the ratio of emitted to absorbed photons. To account for the wavelength-dependent sensitivity of the detector and losses due to absorption by optical components, a stabilized tungsten halogen light source with known spectral power distribution (Thorlabs SLS201/M, 300-2600 nm) was placed in front of the entrance of the integrating sphere and its spectrum recorded in the relevant spectral ranges. In the spectral range around 575 nm, the lamp spectrum was measured while the integrating sphere was empty. In the spectral range of sample emission (900-1365 nm), the lamp spectrum was measured while a cuvette filled with 1 mL of toluene was placed inside the integrating sphere to correct for absorption of SWNT PL by toluene. Correction functions for the two spectral ranges were calculated from the ratio of the measured to the theoretical lamp spectrum and applied to each spectrum before integration. The measurement error is primarily determined by the measurement of laser absorption and is estimated to be 10 %, *i.e.* for a PLQY of 3 % an error of 0.3 % is assumed.

**Measurement of pristine and functionalized SWNTs.** As a reference to the functionalized (6,5) SWNTs, the shear force mixed stock dispersion of pristine (6,5) SWNTs was characterized. Prior to the measurement, dispersions of pristine or functionalized (6,5) SWNTs were diluted to an optical density of 0.2 (per cm path length) at the $E_{11}$ transition by addition of toluene, such that self-absorption is negligible.

**Estimation of exciton population.** Samples were excited at 575 nm ($E_{22}$ transition) by the spectrally-filtered output of a pulsed supercontinuum laser source (Fianium WhiteLase SC400, 20 MHz repetition rate, ~6 ps pulse width). The laser entered the cuvette unfocused, thus the beam cross-section was ~3 mm$^2$. The average laser power was ~300 µW, which corresponds to $4\times10^7$ photons per pulse. We estimate the number of created excitons in the sample *via* the number of absorbed photons in the volume of the cuvette that is directly irradiated by the laser. Subsequent emission and reabsorption cycles are neglected. The measured (6,5) SWNT dispersions had typical optical densities of ~0.06 at the $E_{22}$ transition in a 1 cm path length cuvette, corresponding to absorption of ~13 % of incident photons. Hence, $5.2\times10^6$ excitons are created per pulse. Due to the ~50 ns delay between pulses, they can be treated independently. The nanotube concentration in the dispersion is estimated from the peak molar absorptivity of the $E_{11}$ transition[6] and amounts to $2\times10^{16}$ carbon atoms per cm$^3$. Using the geometrical factor of 88 carbon atoms per nm tube length for (6,5) SWNTs and an average tube length of 1.7 µm for the SWNTs in our shear-mixed samples (see **Figure S11**), a concentration of $1\times10^{11}$ nanotubes per cm$^3$ is obtained. The volume directly irradiated by the laser is ~0.03 cm$^3$, thus the number of nanotubes in this volume is $1.5\times10^9$. Hence, the average number of excitons per nanotube is on the order of 10$^{-3}$. Consequently, the PLQY measurements were conducted in the linear excitation regime, where exciton-exciton annihilation and state-filling effects are absent. This is further supported by near-identical $E_{11}*$ to $E_{11}$ PL intensity ratios when samples are measured in this configuration and under continuous lamp excitation (~1 mW cm$^{-2}$).

**PLQY *vs.* defect density for all substituents. Figure S8** shows the PLQY as a function of defect density for all substituents investigated in this work. The PLQY is further resolved into the main spectral contributions: $E_{11}$ and phonon sideband for pristine SWNTs; $E_{11}$, $E_{11}*$ and $E_{11}*^-$ for functionalized SWNTs. For a summary of the general trends, refer to the discussion of **Figure 4a** in the section "Photoluminescence Quantum Yields and Lifetimes" of the main text. There are slight differences in the PLQY evolution when the substituents are compared. Some of this variation, such as the structure around the maximum $E_{11}*$ PLQY in the 4-bromophenyl series, is likely due to data scattering. Further, the $E_{11}$ PLQY of the 4-methoxyphenyl functionalized (6,5) SWNTs appears to converge to ~0.5 %, whereas it converges to zero for all other substituents. This probably results from a combination of two aspects: Firstly, the 4-methoxy-substituted diazonium salt is particularly unreactive due to its electron-donating substituent and thus, even at the highest diazonium salt concentration tested in this work, the defect density is not high enough to strongly suppress $E_{11}$ emission. Secondly, this defect type exhibits the lowest trap depth and thus, thermal detrapping might lead to non-zero $E_{11}$ PLQY even for very high defect densities.

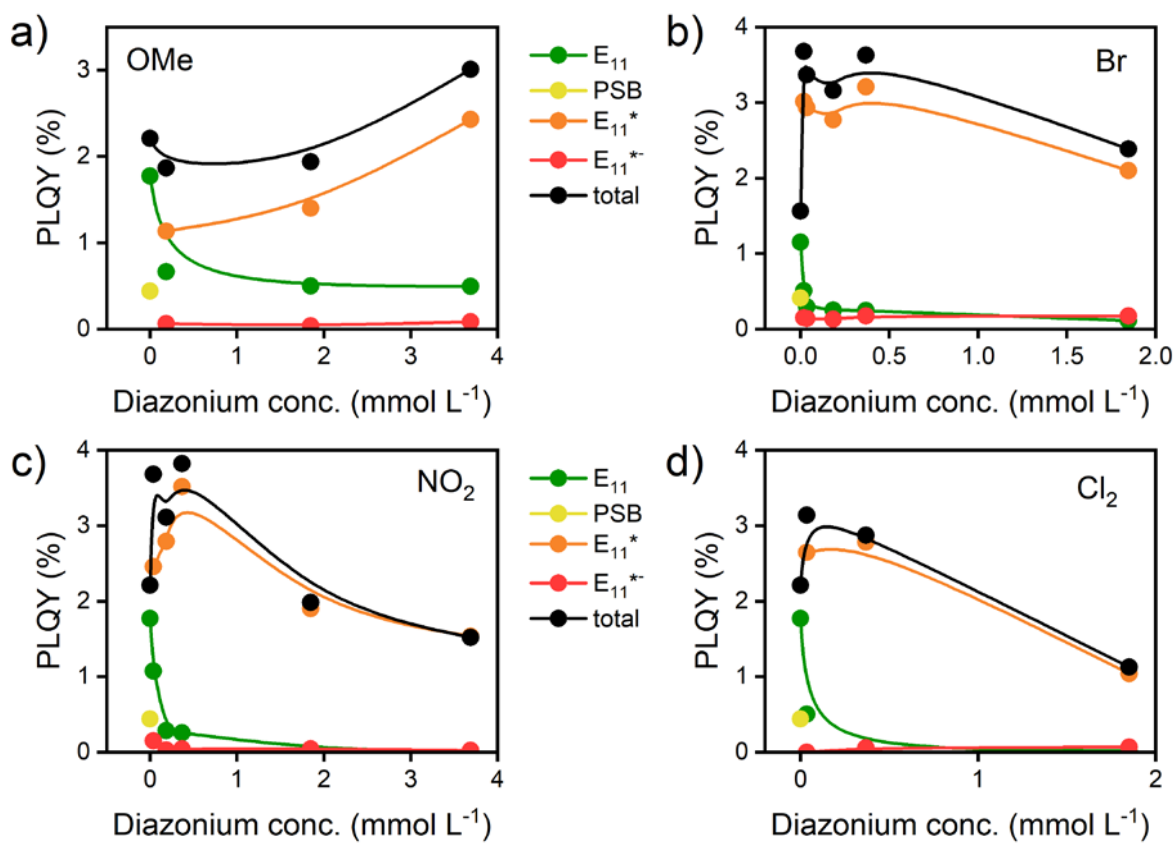

**Figure S8.** Spectral contributions and total PLQY for reference samples of pristine (6,5) SWNTs and (6,5) SWNTs functionalized with different densities of aryl defects bearing different substituents. In the case of pristine SWNTs, the PLQY was split into an $E_{11}$ (900-1080 nm) and a phonon sideband contribution (PSB, 1080-1365 nm). The PLQY of functionalized SWNTs is divided into $E_{11}$ (900-1080 nm), $E_{11}^*$ (1080-1320 nm) and $E_{11}^{*-}$ (1320-1365 nm) contributions. Solid lines are guides to the eye. a) 4-Methoxyphenyl, b) 4-bromophenyl, c) 4-nitrophenyl and d) 3,5-dichlorophenyl defects.

## Defect State Decay Dynamics

The defect state decay dynamics of functionalized (6,5) SWNTs were investigated using time-correlated single photon counting (TCSPC). Dispersions were filled in a quartz cuvette and the excitation laser source (575 nm, ~6 ps pulse width, 20 MHz repetition rate) focused into the liquid *via* a 50× IR-optimized objective (N.A. 0.65, Olympus). Low pulse energy densities of ~4 µJ cm$^{-2}$ were used in all measurements. A spectrograph (Acton SpectraPro SP2358) was used to select the $E_{11}$* emission and this light was focused onto a gated InGaAs/InP avalanche photodiode (Micro Photon Devices) through a 20× IR-optimized objective (Mitutoyo). The instrument response function (IRF) was estimated for each sample from the fast, detector-limited PL decay of the (6,5) SWNTs at the $E_{11}$ transition at 1000 nm. We confirmed that the IRF obtained in this way matches the response obtained from laser light scattering off a silica bead dispersion and has a FWHM of ~84 ps. Taking the IRF into account, all decay curves were fitted to a biexponential model in a reconvolution procedure.

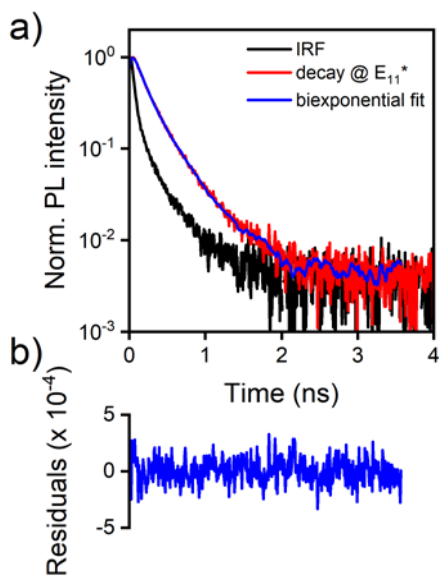

**Figure S9.** a) TCSPC histogram showing the PL decay (red curve) at the $E_{11}$* transition (1160 nm) of 4-bromophenyl-functionalized (6,5) SWNTs in toluene. The biexponential fit (blue curve) was determined *via* a reconvolution procedure considering the IRF (black curve). Note that the roughness of the fit (blue curve) originates from reconvolution with the IRF. b) Residuals of the fit to the decay.

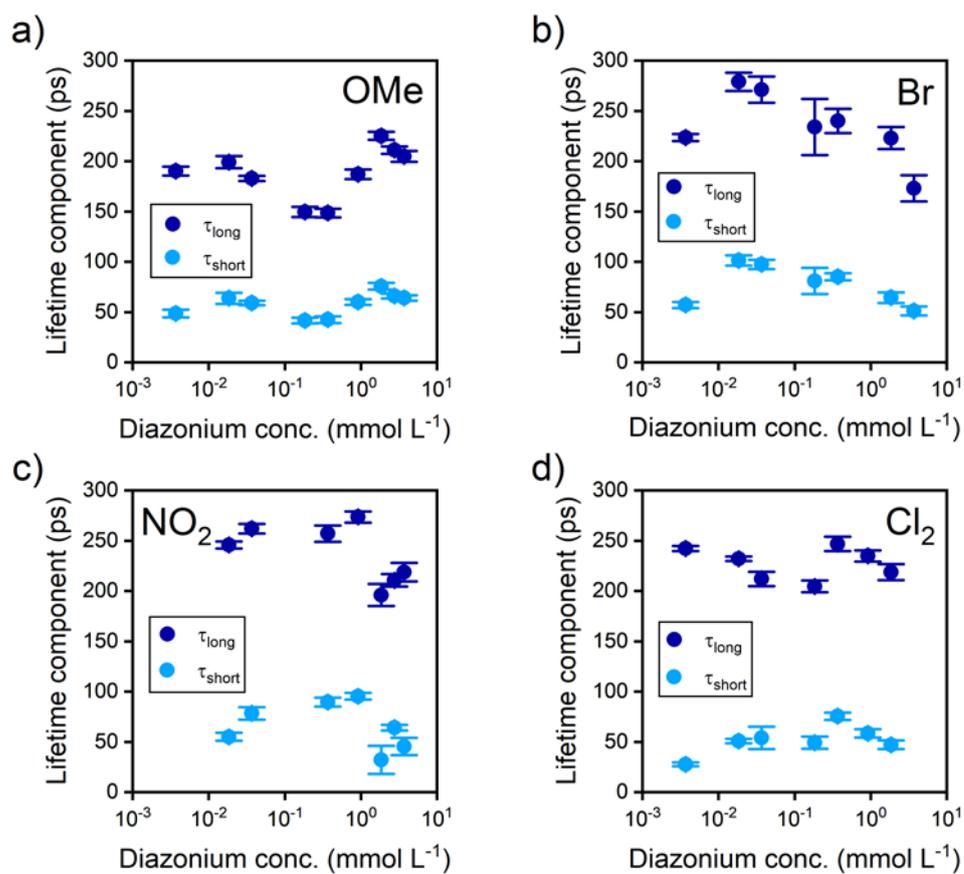

**Figure S10.** Long and short lifetime components as a function of diazonium salt concentration for a) 4-methoxy-, b) 4-bromo-, c) 4-nitro- and d) 3,5-dichlorophenyl defects on (6,5) SWNTs in toluene.

## AFM Statistics of SWNT Length Distributions

Shear mixed and shortened (6,5) SWNTs were spin-coated (2000 rpm, 60 sec) from toluene dispersions ($E_{11}$ absorbance ~0.2 cm$^{-1}$) onto native silicon wafers and excess polymer was rinsed off with tetrahydrofuran and isopropanol. Atomic force micrographs were recorded with a Bruker Dimension Icon atomic force microscope in tapping mode.

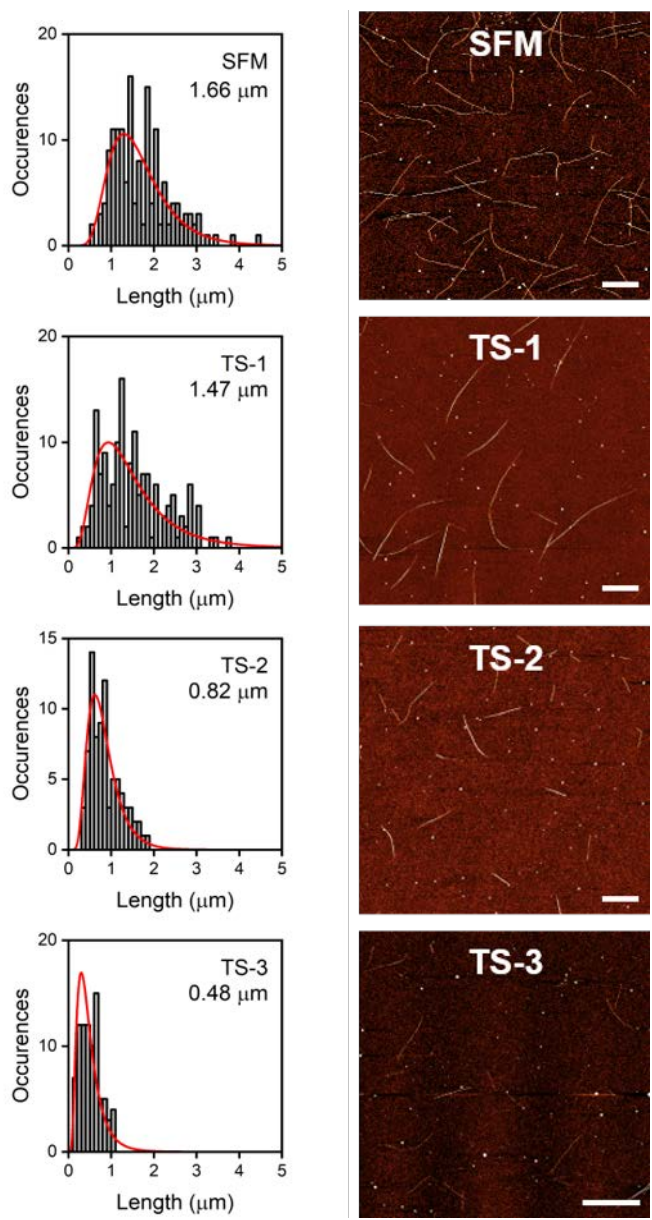

**Figure S11.** Nanotube length histograms and representative tapping mode atomic force micrographs for batches of shear mixed and shortened tubes. All scale bars are 1 µm. SFM (shear force mixed stock), TS-1 (additional 4.5 h tip sonication), TS-2 (additional 12 h tip sonication) and TS-3 (additional 23 h tip sonication).

## Characterization of Printed (6,5) SWNT Films

Pristine (ref) and 4-bromophenyl-functionalized (funct) (6,5) SWNTs were aerosol-jet printed on glass (Aerosol Jet 200 printer, Optomec, with an ultrasonic atomizer). The number of printing cycles was varied (1×, 2×, 4×, 8×) to produce films of different thicknesses as clearly seen in the brightfield optical microscope images (**Figure S12**).

Raman mapping (incident wavelength 532 nm) of the $G^+$ mode intensity was used to confirm that the stripes in one pair have similar thickness. The $G^+$ mode intensity was integrated along the stripes and the resulting Raman intensity profiles are provided in **Figure S12**. The stripes (especially the thickest ones) have very similar thickness.

Homogeneous illumination of the stripes with a 640 nm diode laser excited the (6,5) SWNTs and their photoluminescence was imaged onto a 2D InGaAs camera. The emission from the stripe of functionalized SWNTs is stronger by a factor of ~2 relative to the pristine SWNTs, which is clearly larger than the variation in the density of SWNTs as estimated by Raman microscopy. The brightness of the pixels was corrected for the wavelength-dependent detection efficiency and absorption due to optical components in our setup taking into account the emission spectra of pristine SWNTs and functionalized SWNTs of the respective stripes. We find this correction to be small though (~10 %), so it does not affect the overall conclusion.

As a comparison with the Raman intensity profiles, the PL intensity was integrated along the stripe. Note that a slight signal offset in the region between the stripes is also a result of the correction for the detection efficiency.

To quantify the brightening induced by functionalization, the emission from the stripes was analyzed as a function of film thickness (see **Figure S13**). The relationship between the Raman $G^+$ mode intensity and the number of printing cycles is linear up to a value of 4 cycles, but then saturates for 8 cycles (**Fig. S13a**). This deviation is most likely due to self-absorption of the Raman-scattered light within the thick film. Since the incident laser wavelength is 532 nm and the $G^+$ mode is shifted by 1590 cm$^{-1}$, the scattered light has a wavelength of 581 nm, which matches the $E_{22}$ absorption band in (6,5) SWNTs. Based on the linear correlation between the Raman signal and the number of printing cycles in the low thickness range, it is justified to use the number of printing cycles as a measure of film thickness. The spectrally-integrated PL intensity depends linearly on the number of printing cycles for both the pristine and the functionalized SWNTs (**Fig. S13b**). Finally, the ratio of the slopes of these linear correlations provides a thickness-independent estimate of the PL enhancement induced by $sp^3$ functionalization (**Fig. S13c and S13d**). A brightening factor of 1.7 is computed, in excellent agreement with the absolute PLQY measurements performed on spin-coated films (refer to the main text).

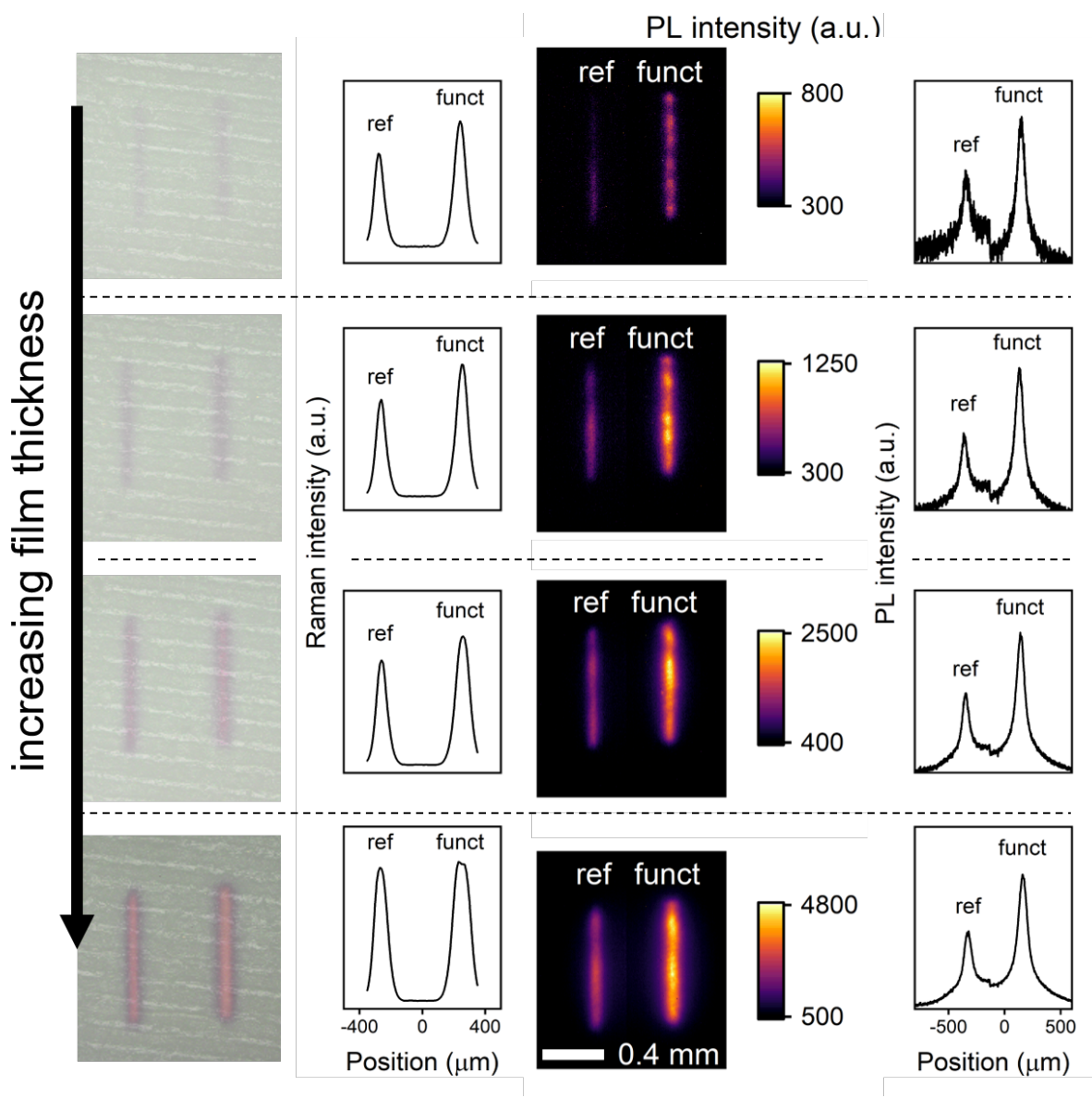

**Figure S12.** From left to right, brightfield optical microscope images, integrated Raman $G^+$ mode intensity profiles, PL micrographs and integrated PL intensity profiles of printed films of pristine (ref) and functionalized (funct) (6,5) SWNTs on glass with different thicknesses.

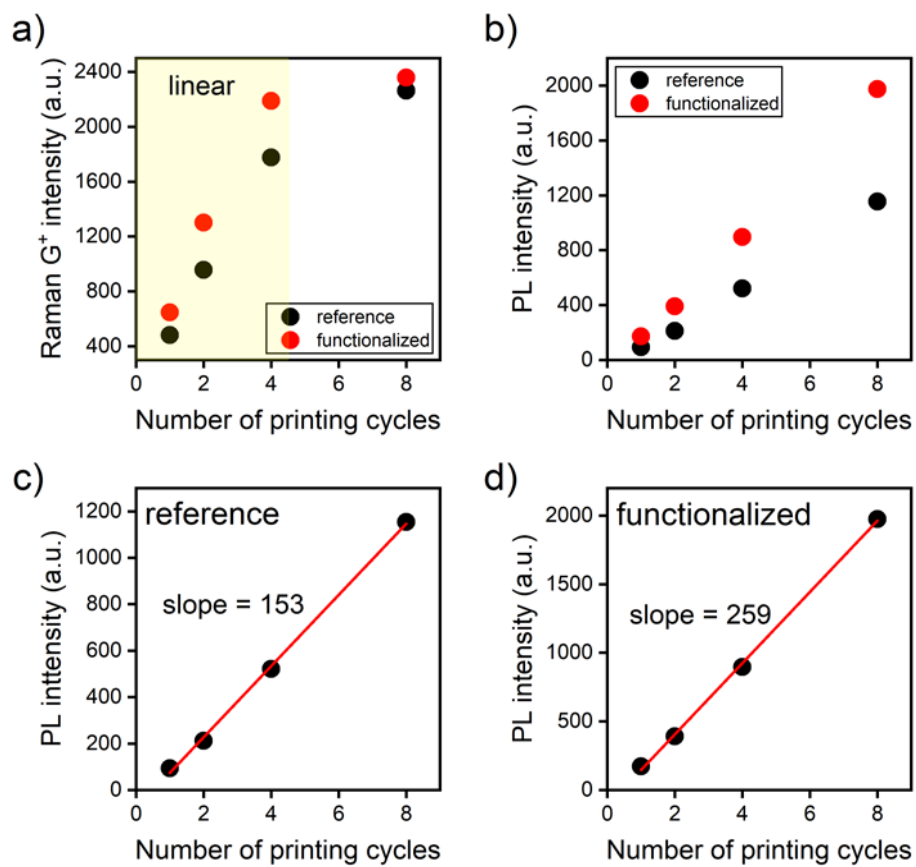

**Figure S13.** a) Raman G$^+$ mode intensity *vs.* number of printing cycles. b) PL intensity *vs.* number of printing cycles. Linear fits to the PL intensity *vs.* number of printing cycles data for films of c) pristine and d) functionalized SWNTs.

# REFERENCES

1.	Gomberg, M.; Bachmann, W. E., The Synthesis of Biaryl Compounds by Means of the Diazo Reaction. *J. Am. Chem. Soc.* **1924,** *46*, 2339-2343.

2.	Schmidt, G.; Gallon, S.; Esnouf, S.; Bourgoin, J. P.; Chenevier, P., Mechanism of the Coupling of Diazonium to Single-Walled Carbon Nanotubes and Its Consequences. *Chem. Eur. J.* **2009,** *15*, 2101-2110.

3.	Wilson, H.; Ripp, S.; Prisbrey, L.; Brown, M. A.; Sharf, T.; Myles, D. J. T.; Blank, K. G.; Minot, E. D., Electrical Monitoring of $sp^3$ Defect Formation in Individual Carbon Nanotubes. *J. Phys. Chem. C* **2016,** *120*, 1971-1976.

4.	Beadle, J. R.; Korzeniowski, S. H.; Rosenberg, D. E.; Garcia-Slanga, B. J.; Gokel, G. W., Phase-Transfer-Catalyzed Gomberg-Bachmann Synthesis of Unsymmetrical Biarenes: A Survey of Catalysts and Substrates. *J. Org. Chem.* **1984,** *49*, 1594-1603.

5.	Piao, Y.; Meany, B.; Powell, L. R.; Valley, N.; Kwon, H.; Schatz, G. C.; Wang, Y., Brightening of Carbon Nanotube Photoluminescence through the Incorporation of $sp^3$ Defects. *Nat. Chem.* **2013,** *5*, 840-845.

6.	Streit, J. K.; Bachilo, S. M.; Ghosh, S.; Lin, C. W.; Weisman, R. B., Directly Measured Optical Absorption Cross Sections for Structure-Selected Single-Walled Carbon Nanotubes. *Nano Lett.* **2014,** *14*, 1530-1536.

7.	Dresselhaus, M. S.; Jorio, A.; Souza Filho, A. G.; Saito, R., Defect Characterization in Graphene and Carbon Nanotubes Using Raman Spectroscopy. *Phil. Trans. R. Soc. A* **2010,** *368*, 5355-5377.

8.	Naumov, A. V.; Ghosh, S.; Tsyboulski, D. A.; Bachilo, S. M.; Weisman, R. B., Analyzing Absorption Backgrounds in Single-Walled Carbon Nanotube Spectra. *ACS Nano* **2011,** *5*, 1639-1648.

9.	Pfohl, M.; Tune, D. D.; Graf, A.; Zaumseil, J.; Krupke, R.; Flavel, B. S., Fitting Single-Walled Carbon Nanotube Optical Spectra. *ACS Omega* **2017,** *2*, 1163-1171.

10.	Graf, A.; Zakharko, Y.; Schießl, S. P.; Backes, C.; Pfohl, M.; Flavel, B. S.; Zaumseil, J., Large Scale, Selective Dispersion of Long Single-Walled Carbon Nanotubes with High Photoluminescence Quantum Yield by Shear Force Mixing. *Carbon* **2016,** *105*, 593-599.